\documentclass[3p,12pt]{elsarticle}
\usepackage{lineno,hyperref}
\usepackage{amsmath,bm,geometry}
\usepackage{amsfonts}
\usepackage{color}
\usepackage{nomencl,etoolbox,ragged2e,siunitx}
\usepackage{mathtools}
\usepackage{multirow}
\usepackage{caption}
\usepackage{parskip}
\usepackage{calligra}
\usepackage{makeidx}
\usepackage{tikz}
\usepackage{subcaption}
\usepackage[utf8]{inputenc}
\usepackage[english]{babel}
\usepackage{amsthm}

\usetikzlibrary{matrix}
\makeindex
\DeclareMathAlphabet{\mathcalligra}{T1}{calligra}{m}{n}
\newcommand*{\defeq}{\stackrel{\text{def}}{=}}
\begin{document}
	
	\begin{frontmatter}
		
		\title{Modeling and computation for non-equilibrium gas dynamics: beyond kinetic relaxation model}
		
		\author[hkust]{Xiaocong Xu}
		\ead{xxuay@connect.ust.hk}
		
		\author[hkust]{Yipei Chen}
		\ead{ychendh@connect.ust.hk}
		
		\author[hkust]{Kun Xu\corref{cor1}}
		\ead{makxu@ust.hk}

		\cortext[cor1]{Corresponding author}
		\address[hkust]{Mathematics Department  \\
			Hong Kong University of Science and Technology  \\
			Clear Water Bay, Kowloon, Hong Kong}

		\begin{abstract}
			
The non-equilibrium gas dynamics is described by the Boltzmann equation, which can be solved numerically through the deterministic and stochastic methods. The deterministic Boltzmann solver is usually based on the discrete velocity method (DVM) in the particle velocity space and the numerical partial differential equation discretization in the physical space. The stochastic direct simulation Monte Carlo (DSMC) method is a direct modeling of the Boltzmann process through particle transport and collision. Due to the complicated collision term of the Boltzmann equation, many kinetic relaxation models have been proposed and used in the past seventy years for the study of rarefied flow. In order to develop a multiscale method for the rarefied and continuum flow simulation, by adopting the integral solution of the kinetic model equation a DVM-type unified gas-kinetic scheme (UGKS) has been constructed. The UGKS models the gas dynamics on the cell size and time step scales while the accumulating effect from particle transport and collision has been taken into account within a time step. Under the UGKS framework, a unified gas-kinetic wave-particle (UGKWP) method has been further developed for non-equilibrium flow simulation, where the time evolution of gas distribution function is composed of analytical wave and individual particle. In the highly rarefied regime, particle transport and collision will play a dominant role. Due to the single relaxation time model for particle collision, there is a noticeable discrepancy between the UGKWP solution and the full Boltzmann or DSMC result, especially in the high Mach and Knudsen number cases. In this paper, besides the kinetic relaxation model, a modification of particle collision time according to the particle velocity will be implemented in UGKWP. As a result, the new model greatly improves the performance of UGKWP in the capturing of non-equilibrium flow. There is a perfect match between UGKWP and DSMC or Boltzmann solution in the highly rarefied regime. In the near continuum and continuum flow regime, the UGKWP will gradually get back to the macroscopic variables based Navier-Stokes flow solver at small cell Knudsen number.

		\end{abstract}
		
		
	\end{frontmatter}

	\section{Introduction}

	According to the Knudsen number, defined as the ratio of the particle mean free path over a characteristic
	length scale, the flow regime can be classified into continuum, transition, and rarefied  one.
	In many engineering applications, such as hypersonic flow around a vehicle in near space flight, multiple regimes can co-exist in a
	single computation, where the local Knudsen number can vary widely with differences on many orders of magnitude \cite{jiang}.
	The development of multiscale method for all flow regimes becomes necessary \cite{xu-book}.
	The Boltzmann equation describes gas dynamics on the kinetic particle mean free path and collision time scale \cite{chapman}.
	Many Boltzmann solvers based on the discrete velocity method (DVM) have been constructed
	with the requirement on the cell size and time step at the kinetic scale \cite{chu,yang,li,mieussens,aristov}.
	Instead of solving the complicated collision term of the Boltzmann equation, the direct simulation Monte Carlo (DSMC) mimics the Botzmann transport and collision process using stochastical particles and achieves great success for the rarefied flow study \cite{bird}.
	Same as direct Boltzmann solver, the DSMC also requires the cell size and time step to be less than the particle mean free path and
	collision time.
	In the study of rarefied flow, instead of the full Boltzmann collision term, many kinetic relaxation models have been
	proposed in the past seventy years \cite{bgk, holway, shakhov, rykov}.
	
	Based on the gas-kinetic relaxation models and DVM discretization,  a unified gas-kinetic scheme (UGKS) has been
	developed for the rarefied and continuum flow study \cite{xu-huang}. The evolution of the gas distribution function in UGKS is
	based on the integral solution of the kinetic model equation, where the accumulating effect from particle transport and collision within a numerical time step is used in the construction of the multiscale method.
	The integral solution is basically composed of the hydrodynamic evolution from  the
	integration of equilibrium state and the kinetic particle free transport from the initial non-equilibrium gas distribution.
	The weights between these two scales dynamics depend on the cell's Knudsen number. Therefore, the integral solution can give an
	accurate representation for both continuum and free molecule flow and used in the construction of the multiscale scheme.
	In the continuum flow regime at high Reynolds number, the UGKS gets back to the standard Navier-Stokes flow solver with a numerical time step
	being  much larger than the particle collision time \cite{xu-book}.
	In order to improve the efficiency and reduce the memory cost, the DVM-based gas distribution function in UGKS is replaced by a combination of
	particle and wave in the newly developed unified gas-kinetic wave-particle (UGKWP) method \cite{liu,zhu,xu-c}.
	The evolution of particle and wave in UGKWP is still controlled by the integral solution of the kinetic relaxation model.
	Even though, accurate solutions can be obtained from UGKWP in all flow regimes, especially in the near continuum and continuum regimes,
	there are still some discrepancies in the solution in the highly non-equilibrium flow regime, such as the temperature distribution inside shock layer at high Mach number. The reason for the difference is mainly coming from the single relaxation time for all speed particles
	in the kinetic model equation, while the physical relaxation time varies in the Boltzmann collision term and the DSMC according to particle velocity. In fact, according to basic kinetic theory, the particle collision time $\tau$ is related to the mean free path $\ell$ and particle velocity $|\vec{v}|$, such as $\tau = \ell/|\vec{v}|$. Due to the wave-particle decomposition in UGKWP, in the highly rarefied regime,
	the flow evolution in UGKWP is mainly controlled by particles. Therefore, the collision time of particle transport in UGKWP can be simply modified according to particle velocity. Since this only effects the free transport distance of individual particle, the mass, momentum, and energy of the system are fully conserved.  This direct modeling on the particle transport greatly improves the performance of UGKWP for the
	rarefied flow solution, where excellent results can be obtained and matched with the  Boltzmann or DSMC solutions.
	As a result, through direct modeling on the particle collision time the dynamics in the above UGKWP is beyond the single relaxation kinetic model equation. Multiple relaxation time according to particle velocity can be directly implemented in the scheme.
	In the near continuum regime, the particles in UGKWP will disappear and a continuum Navier-Stokes flow solver is recovered.
	
	This paper is arranged as follows. Section 2 reviews the kinetic model equation based multiscale
	UGKWP method and presents the modification on the particle collision time according to particle velocity.
	Section 3 tests the UGKWP in shock structure calculations for monatomic and diatomic gases and flow passing through a cylinder
    at high Mach numbers.
	The last section is the conclusion.

	\section{Kinetic relaxation model and unified gas-kinetic wave-particle method}
	The kinetic relaxation model for the evolution of gas distribution function $f$ is
	\begin{equation}\label{simplefiedRykov}
	\frac{\partial f}{\partial t}+ \vec{v}\cdot\nabla_{\vec{x}} f = \frac{M-f}{\tau},
	\end{equation}
	where $M$ is the equilibrium state, $\vec{v}$ is the particle velocity, and $\tau$ is the relaxation or collision time.
	For monatomic and diatomic gases, the BGK, ES, Shakhov, and Rykov models can be formulated in the above relaxation
	form \cite{bgk, holway, shakhov, rykov}.
	For a local constant collision time $\tau$, the integral solution of Eq.\eqref{simplefiedRykov} can be written as \cite{xu,xu-huang},
	\begin{equation}\label{integral-solution}
	f(\vec{x},t,\vec{v},\vec{\xi})=\frac{1}{\tau}\int_{0}^t e^{-(t-t')/\tau} M(\vec{x}',t',\vec{v},\vec{\xi}) dt'+e^{-t/\tau}f_0(\vec{x}-\vec{v}t),
	\end{equation}
	where $\xi$ is the internal variable and $f_0$ is the initial gas distribution function at $t=0$.
	In both equilibrium state and initial distribution, the particle transports along characteristics $\vec{x}'=\vec{x}+\vec{v}(t'-t)$.
	
	The UGKWP is constructed on a discretized physical space $\sum_i \Omega_i\subset\mathcal{R}^3$ and discretized time $t^n\in\mathcal{R}^+$
	\cite{liu,zhu,xu-c}.
	Initially, at the beginning of each time step, the distributions of macroscopic flow variables and microscopic gas distribution function
inside each control volume are known.
	The cell averaged conservative flow variables, such as $\vec{W}_i=(\rho_i,\rho_i\vec{U}_i,\rho_iE_i)$ in a physical control volume $\Omega_i$, are defined as
	\begin{equation*}
	\vec{W}_i^n=\frac{1}{|\Omega_i|}\int_{\Omega_i}\vec{W}^n (\vec{x}) d\vec{x},
	\end{equation*}
	and the corresponding distribution of particles is given by $f_0$, with the condition
 $$\vec{W}_i^n = \int f_0 \psi d\Xi,$$
 where $\psi = (1, \vec{v}, \frac{1}{2} (\vec{v}^2 +\vec{\xi}^2))^T$ is the vector of moments for the conservative flow variables and $d\Xi =d\vec{v}d\vec{\xi}$.

	The UGKWP will update both the macroscopic flow variables and the gas distribution function.
The macroscopic variables will be updated in a conservative form
	\begin{equation}\label{update-w}
	\vec{W}_i^{n+1}=\vec{W}_i^{n}-\frac{\Delta t}{|\Omega_i|} \sum_{l_s\in\partial\Omega_i} |l_s|\vec{F}_{s} ,
	\end{equation}
	where $l_s \in\partial\Omega_i$ is the cell interface with center $\vec{x}_s$ and outer unit normal vector $\vec{n}_s$. The flux function for the macroscopic variables at the cell interfaces are constructed from Eq. (\ref{integral-solution}),
	\begin{equation}\label{flux-w-ugks}
	\begin{aligned}
	\vec{F}_{s} &= \frac{1}{\Delta t}\int_{0}^{\Delta t}\int f(\vec{x}_s,t,\vec{v},\vec{\xi})\vec{v}\cdot \vec{n}_s \vec{\psi} d\Xi dt \\
	&=\frac{1}{\Delta t}\int_{0}^{\Delta t}\int\bigg[\frac1\tau\int_0^{t} e^{(t'-t)/\tau}M(\vec{x}'_s,t',\vec{v},\vec{\xi})dt'+
	e^{-t/\tau}f_0(\vec{x}_s-\vec{v}t)\bigg] \vec{v}\cdot \vec{n}_s \vec{\psi} d\Xi dt,
	\end{aligned}
	\end{equation}
	with the characteristic line $\vec{x}'_s=\vec{x}_s+\vec{v}(t'-t)$.
	The equilibrium flux from the integration of Maxwellian distribution is denoted as $\vec{F}_{eq,s}$,
	\begin{equation}\label{Fg}
	\vec{F}_{eq,s}\defeq\frac{1}{\Delta t}\int_{0}^{\Delta t}\int\frac1\tau\int_0^{t} e^{(t'-t)/\tau}M(\vec{x}'_s,t',\vec{v}, \vec{\xi})dt' \vec{v}\cdot\vec{n}_s\vec{\psi} d\Xi dt,
	\end{equation}
	which can be evaluated analytically from the distribution of macroscopic flow variables $\vec{W}^n$ through the gas-kinetic formulation \cite{xu}. The flux coming from the free transport of $f_0$ is given by $\vec{F}_{fr,s}$,
	\begin{equation}\label{Ff}
	\vec{F}_{fr,s}\defeq\frac{1}{\Delta t}\int_{0}^{\Delta t}\int e^{-t/\tau} f_0(\vec{x}_s-\vec{v}t) \vec{v}\cdot \vec{n}_s\vec{\psi} d\Xi dt,
	\end{equation}
	which is evaluated by tracking the particles from the initial distribution $f_0$ \cite{liu,xu-c}.
Specifically, the updates of macroscopic variables will become,
	\begin{equation} \label{w}
	\vec{W}^{n+1}_i=\vec{W}^n_i - \frac{\Delta t}{|\Omega_i|} \sum_{l_s\in\partial \Omega_i}|l_s| \vec{F}_{eq,s} +  \frac{\Delta t}{|\Omega_i|}\vec{W}_{fr,i} ,
	\end{equation}
	where $\vec{W}_{fr,i}$	is the net free streaming flow of cell $i$ calculated by counting the particles passing through the cell interface during a time step.
One of the outstanding features of UGKWP is that the gas distribution function $f_0$ is composed of collisional particle and
collisionless particle in the evolution process within a time step $\Delta t$. The flux in $\vec{W}_{fr,i}$ from the collisional particle
can be evaluated analytically. The collisionless particle number reduces as $\exp{-\Delta t/ \tau}$. In the continuum flow regime, all fluxes from
$f_0$ can be evaluated analytically and the UGKWP will become a standard NS solver for the update $\vec{W}$ only. The detailed formulation is
given next.

	The evolution of particle $P_k(m_k,\vec{x}_k,\vec{v}_k, e_k, t_{f,k})$ is represented by its mass $m_k$,
	position coordinate $\vec{x}_k$, velocity coordinate $\vec{v}_k$, internal energy $e_k$, and free streaming time $t_{f,k}$.
	The particle evolution follows the same integral solution of the kinetic model equation,
	\begin{equation}\label{integral-solution1}
	f(\vec{x},t,\vec{v},\vec{\xi})=\frac{1}{\tau}\int_{0}^t e^{-(t-t')/\tau} M(\vec{x}',t',\vec{v},\vec{\xi}) dt'+e^{-t/\tau}f_0(\vec{x}-\vec{v}t),
	\end{equation}
	which results in 	
	\begin{equation}\label{particle}
	f(\vec{x},t,\vec{v},\vec{\xi})=(1-e^{-t/\tau})M^+(\vec{x},t,\vec{v},\vec{\xi})+e^{-t/\tau}f_0(\vec{x}-\vec{v}t).
	\end{equation}
	The above $M^+$ is named as the hydrodynamic distribution function with analytical formulation, i.e., the wave formulation of the gas distribution function.
	The initial particle distribution $f_0$  has a probability of $e^{-t/\tau}$ to free streaming and a probability of $(1-e^{-t/\tau})$ to colliding with other particle, and the post-collision distribution follows the distribution $M^+(\vec{x},t,\vec{v},\vec{\xi})$.
	The time for the free streaming to stop and follow the distribution $M^+$ is called the first collision time $t_c$.
	The cumulative distribution function of the first collision time is
	\begin{equation}\label{tc-distribution}
	F(t_c<t)=1-\exp(-t/\tau),
	\end{equation}
	from which $t_c$ can be sampled as $t_c=-\tau\ln(\eta)$ with $\eta$ generated from a uniform distribution $U(0,1)$. For a particle $P_k$, the free streaming time is given by,
	\begin{equation}\label{freetime}
	t_{f,k} =
	\begin{cases}
	-\tau \ln (\eta)&\mbox{if $	-\tau \ln (\eta) < \Delta t$} ,\\
	\Delta t &\mbox{if $	-\tau \ln (\eta) > \Delta t$, }
	\end{cases}
	\end{equation}
	where $\Delta t$ is the time step. In a numerical time step from $t^{n}$ to $t^{n+1}$, all simulating particles in UGKWP method can be categorized into two groups: the \textbf{collisionless particle} \index{collisionless particle} $P^f$ and the \textbf{collisional particle} \index{collisional particle} $P^c$.
	The categorization is based on the relation between the free streaming time $t_{f}$ and the time step $\Delta t$.
	More specifically, the collisionless particle is defined as the particle whose free streaming time $t_f$ being greater than or equal to the time step $\Delta t$,
	and the collisional particle is defined as the particle whose free streaming time $t_f$ being smaller than $\Delta t$.
	For the collisionless particle, its trajectory is fully tracked during the whole time step.
	For collisional particle, the particle trajectory is tracked till $t_f$.
	Then the particle's mass, momentum, and energy are merged into the macroscopic quantities in that cell and the simulation particle gets eliminated.
	Those eliminated particles may  get re-sampled once the updated total macroscopic quantities $\vec{W}^{n+1}$ and the fraction from the
	eliminated particles $\vec{W}^{n+1, h}$ are obtained.
	
	Now the particle will take free streaming for a period of $t_{f,k}$,
	\begin{equation}\label{stream}
	\vec{x}_{k}^{n+1} = \vec{x}_{k}^{n} + \vec{v}_k t_{f,k}.
	\end{equation}
	The net free streaming flow of cell $i$ within a time step $\Delta t$ can be calculated by counting the particles passing through the cell interface, which can be written as,
	\begin{equation}\label{particleevo}
	\vec{W}_{fr,i}=\frac{1}{\Delta t}\left(  \sum_{k\in P_{\partial \Omega_i^{+}}} \vec{W}_{P_k} - \sum_{k\in P_{\partial \Omega_i^{-}}} \vec{W}_{P_k}\right),
	\end{equation}
	where $\vec{W}_{P_k}=\left( \omega_km_k,\omega_km_k \vec{v}_k,\frac12 m_k \left(\omega_k\vec{v}_k^2+\kappa _ke_k \right) \right)^T$,
	$P_{\partial \Omega_i^{+}}$ is the index set of the particles streaming into cell $\Omega_i$ during a time step, and $P_{\partial \Omega_i^{-}}$ is the index set of the particles streaming out of cell $\Omega_i$.
	Then, the update of total macroscopic flow variables in Eq.\eqref{w} is fully determined.
	The update of hydrodynamic flow variables originated from the eliminated particles are
	$$\vec{W}_i^{n+1,h} = \vec{W}_i^{n+1} -\vec{W}_{i}^{n+1,p} ,$$
	where $\vec{W}_{i}^{n+1,p}$ is macroscopic flow variables of all remaining particles in the cell $i$, which can be evaluated by
	adding their mass, momentum, and energy together.

	Theoretically, we can go to the next time step calculation by sampling particles from the distribution $M^+$
    with a total mass, momentum, and energy of $\vec{W}^{n+1,h}$, and get the all particle representation of $f_0$ with the
    condition $\vec{W}_i^{n+1} = \int f_0 \psi d\Xi$ again for the next time step evolution. This is so-called unified gas-kinetic particle (UGKP) method \cite{liu,zhu}.
	The new initial distribution $f_0$ in cell $i$ is composed of the newly sampled particles and remaining particles from the
	previous time step.
	However, the UGKWP is further developed from UGKP in the following.
	Not all particles from $\vec{W}_i^{n+1,h}$ need to be sampled, because some sampled particles will become collisional hydro-particles in the next time step from $t^{n+1} $ to $t^{n+2}$ and disappear by collision.
	Therefore, only collisionless hydro-particles from $\vec{W}_i^{n+1,h}$ need to be sampled with $t_{f,k} =\Delta t$.
	Based on the cumulative distribution function of the first collision time in Eq.\eqref{tc-distribution},
	the collisionless hydro-particles sampled from $M^+(\vec{W}^{n+1}_i)$ take a total mass
	of $e^{-\Delta t/\tau_i}|\Omega_i| \rho^{n+1,h}_i$.
The collisional hydro-particles has a total mass $(1- e^{-\Delta t/\tau_i})|\Omega_i| \rho^{n+1,h}_i$ with a corresponding distribution
$M^+$. The numerical flux contribution in $f_0$ from collisional hydro-particle can be evaluated analytically, which is denoted as $\vec{F}_{fr,s}^h$ \cite{liu}.
	Therefore, the flux contribution from $f_0$ in Eq.(\ref{Ff}) can be evaluated as
	\begin{equation*}
	\sum_s |l_s| \vec{F}_{fr,s}=\sum_s |l_s| \vec{F}_{fr,s}^h+\vec{W}_{fr,i}^p,
	\end{equation*}
	where $\vec{W}_{fr,i}^p$ is the flux contribution of all remaining particles from previous time step and the newly sampled collisionless particles,
	\begin{equation*}
	\vec{W}_{fr,i}^p=\frac{1}{\Delta t}\left(  \sum_{k\in P_{\partial \Omega_i^{+}}} \vec{W}_{P_k} - \sum_{k\in P_{\partial \Omega_i^{-}}} \vec{W}_{P_k}\right).
	\end{equation*}
So, in UGKWP the evolution of macroscopic flow variables in Eq.(\ref{w}) becomes
	\begin{equation}\label{updateWP}
	\vec{W}_i^{n+1}=\vec{W}_i^n+\frac{\Delta t}{|\Omega_i|} \left(\sum_{l_s\in\partial \Omega_i}|l_s| \vec{F}_{eq,s}+\sum_{l_s\in\partial \Omega_i}|l_s| \vec{F}_{fr,s}^h + \vec{W}_{fr,i}^p\right).
	\end{equation}

In UGKWP, the number of particles used in evolution depends on the cell Knudsen number $Kn_c = \tau/\Delta t$ and
takes a fraction of macroscopic variables  $e^{-1/Kn_c} \vec{W}$ inside each control volume.
	In the continuum flow regime with $\Delta t \gg \tau$, the particle will gradually disappear and the UGKWP will become a gas-kinteic
	scheme for the Navier-Stokes equations \cite{xu}.
	
	The above UGKWP is based on the single relaxation kinetic model. Even though the viscosity and heat conduction coefficients can be
	correctly defined through the Shakhov or Rykov models, all particles have the same relaxation time $\tau$ which is used
in the determination of free steaming time of the particle in UGKWP.
	In physical reality, the particle mean free path $\ell$ has a clear definition, such as the hard sphere molecule \cite{bird},
	$$\ell = \frac{16}{5} \left( \frac{m}{2\pi k T}    \right)^{1/2} \frac{\mu}{\rho} ,$$
	where $\mu$ is the dynamic viscosity coefficient, $\rho$ is the density, and $T$ is the temperature. The particle collision time is related
	to the particle velocity $|\vec{v}|$, such as $\ell/|\vec{v}|$.
	In order to incorporate this physical reality, a direct modeling on the modification of particle transport in UGKWP is
to get a more reliable	particle collision time for those particles with a relative high velocity. The newly modeled $\tau^*$ has the form
	\begin{equation}
	\tau_* = \left\{\begin{array}{ll}
	\tau & \text { if $|\vec{v} - \vec{U}| \leq 5 \sigma$ } \\
	\frac{1}{1 + 0.1 * |\vec{v} - \vec{U}|/  \sigma} \tau& \text { if $|\vec{v} - \vec{U}| > 5 \sigma$ }
	\end{array}\right.
	\end{equation}
	where $\sigma = \sqrt{RT}$.
	For the particle, the free streaming time is determined by
	$t_f = -\tau_{*} \ln (\eta) $.
For high speed particle, the relative collision time will be reduced according to the particle velocity.
For other particles, the free steaming time remains the same, which has been taken into account properly by the kinetic relaxation model,
	such as recovering the viscosity and heat conduction coefficients.
The above modification of particle collision time will not effect the conservation of the scheme. For a specific particle, its mass, momentum, and energy will remain the same, and the only difference is about the distance it travels.
	The modification of particle free streaming time can be done directly in the UGKWP.
	For the DVM-based UGKS, it is very hard to modify the collision time at particular particle velocity and keep the conservation.
All examples in the
	next section are computed with the above modified UGKWP-$\tau_*$.
The idea of $\tau_*$ modification in UGKWP has the physical similarity with the previous effort of generalizing the Chapman-Enskog expansion
for non-equilibrium flow study \cite{xu-tau}.
	
	\section{Numerical Examples}
In this section, the modified UGKWP will be tested in both 1D shock structure and 2D flow passing through cylinder case at different Mach and Knudsen numbers.
	One of the simplest and most fundamental non-equilibrium gas dynamic phenomena that can be used
	for the model validation is the internal structure of a normal shock wave. There are mainly two reasons
	for this. First, the shock wave represents a flow condition that is far from thermodynamic equilibrium.
	Second, shock wave phenomena is unique in that it allows one to separate fluid dynamics from the boundary condition.
	The boundary condition for a shock wave is simply determined by the Rankine–Hugoniot relation.
	Thus, in the study of shock structure, one is able to identify the flow physics of the modeling scheme.
	Since 1950s, the computation of shock structure has continuously played an important and critical role
	in validating kinetic theory and numerical scheme in  the non-equilibrium flow study \cite{bird70,hu}.

	In order to validate the UGKWP modeling, the shock  structures at different Mach numbers
for monatomic and diatomic gases will be calculated.  Besides the density and temperature distributions, the stress and heat flux will also be presented in some cases and compared with the reference solutions of the Boltzmann equation and DSMC.
	
	In the following calculations, the viscous coefficient is given by,
	\begin{equation}\label{vhs-vs1}
	\mu=\mu_{ref}\left(\frac{T}{T_0}\right)^{\omega},
	\end{equation}
	with the reference viscosity
	\begin{equation}\label{vhs-vs2}
	\mu_{ref}=\frac{15\sqrt{\pi}}{2(5-2\omega)(7-2\omega)}\frac{ \ell}{L},
	\end{equation}
	where $L$ is the characteristic length, $\omega$ is the temperature dependence index.

\begin{figure}[htbp]
		\centering
		\includegraphics[width=0.48\textwidth]{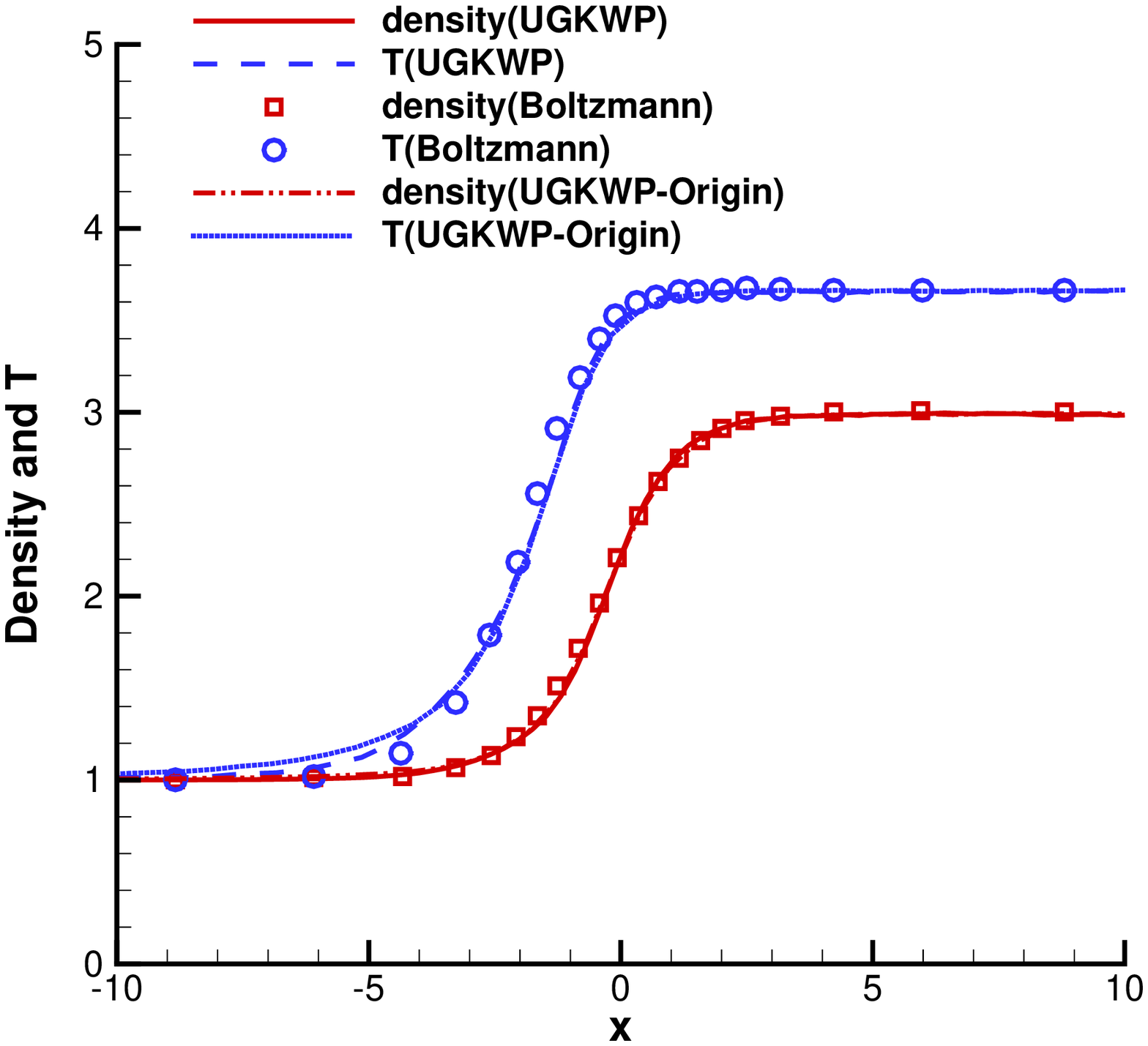}{a}
		\includegraphics[width=0.48\textwidth]{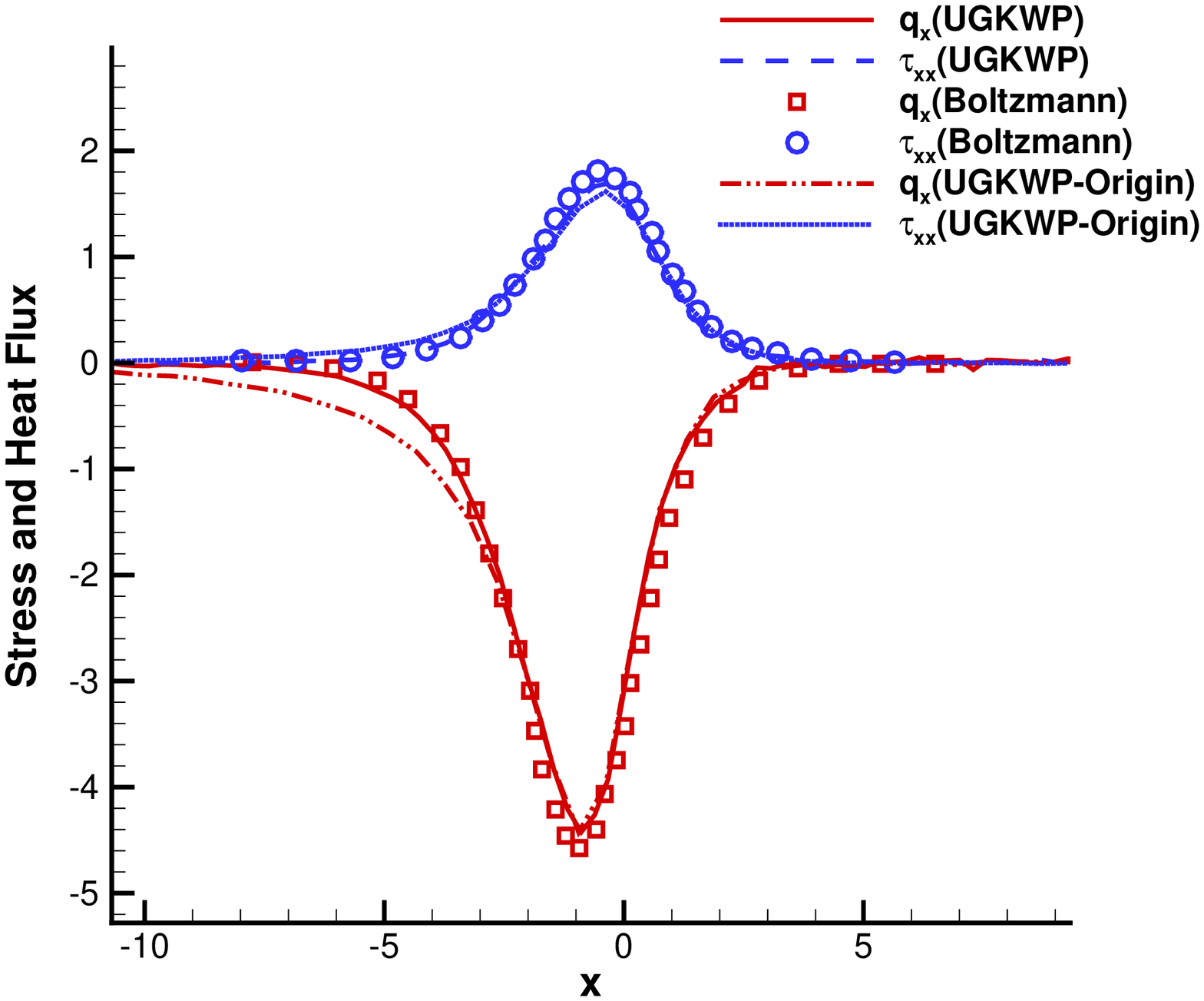}{b}
		\caption{$\mathrm{M} = 3$ argon shock structure calculated by the original UGKWP, updated UGKWP, and the direct Boltzmann solver. The x-coordinate is normalized by $\sqrt{\pi}\ell/2$.}
		\label{shockbolt}
	\end{figure}

	First, the shock structure of monatomic gas will be presented.
Comparisons of UGKWP results are made with the solution of the full Boltzmann equation  for the hard sphere molecules at Mach
	number $\mathrm{M} = 3$ \cite{ohwada}. For the hard sphere molecule, $\omega$ is set to be $0.5$. Fig. \ref{shockbolt} shows the density, temperature, stress and heat flux inside a shock layer computed by the updated and original UGKWP, where better agreement has been obtained between the results from the updated UGKWP and direct Boltzmann solver.

%	For the Mach 8 argon shock structure, Bird’s\cite{bird} DSMC solution with $\mu \sim T^{0.68}$ gave a good agreement with the experiment data. With the same temperature dependency index $\omega=0.68$, we compare the results from the UGKWP and the DSMC method. Fig. \ref{shockdsmc8} shows the solution of density, temperature, heat flux and stresses from the UGKWP and that from DSMC solutions.
	
%	\begin{figure}[htbp]
%		\centering
%		\includegraphics[width=0.48\textwidth]{shock8}{a}
%		\includegraphics[width=0.48\textwidth]{shock8q}{b}
%		\includegraphics[width=0.48\textwidth]{shock8tau}{b}
%		\caption{$\mathrm{M} = 8$ argon shock structure calculated by the UGKWP and the DSMC. The x-coordinate is normalized by $\ell$.}
%		\label{shockdsmc8}
%	\end{figure}
	
\begin{figure}[htbp]
		\centering
		\includegraphics[width=0.48\textwidth]{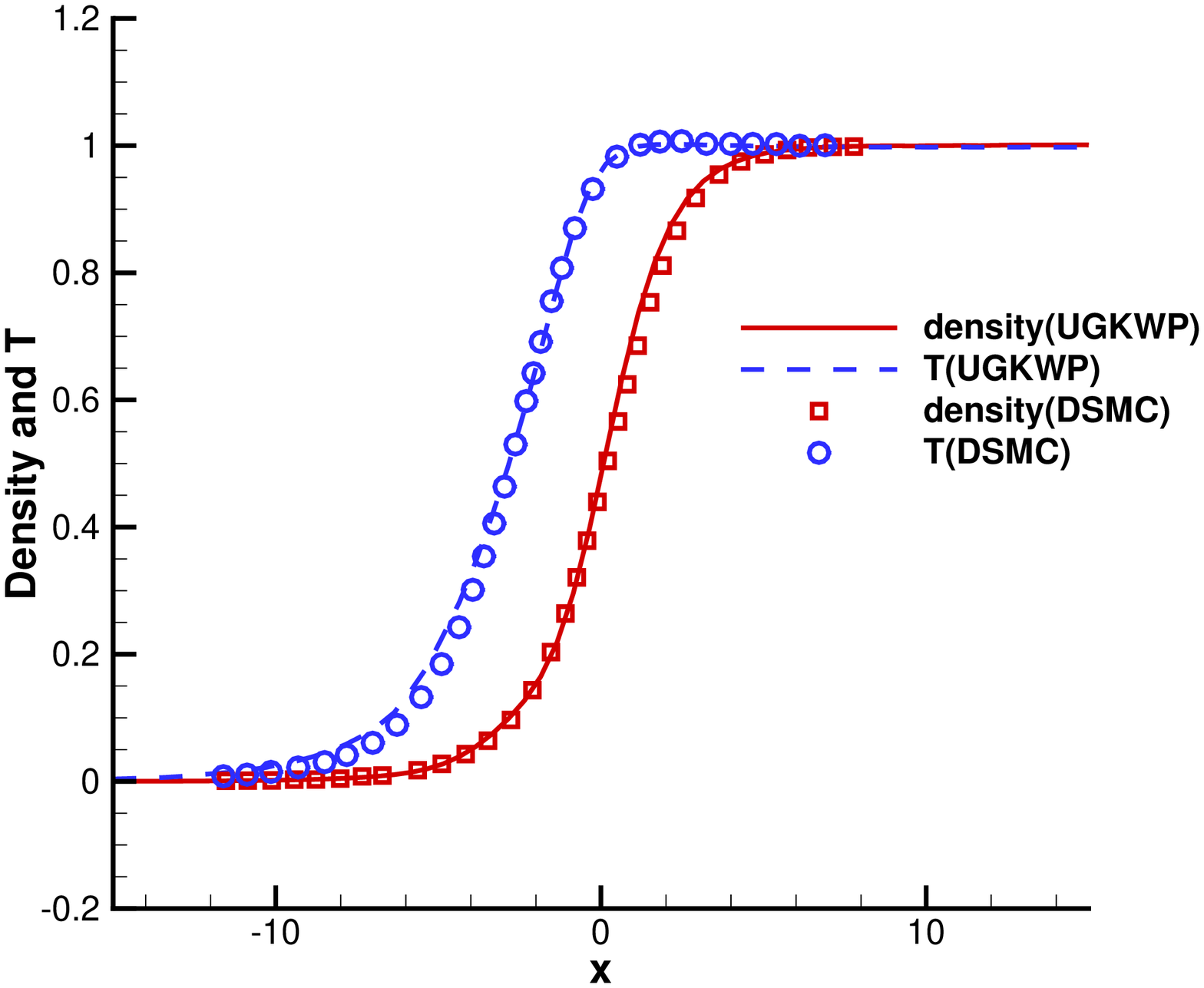}{a}
		\includegraphics[width=0.48\textwidth]{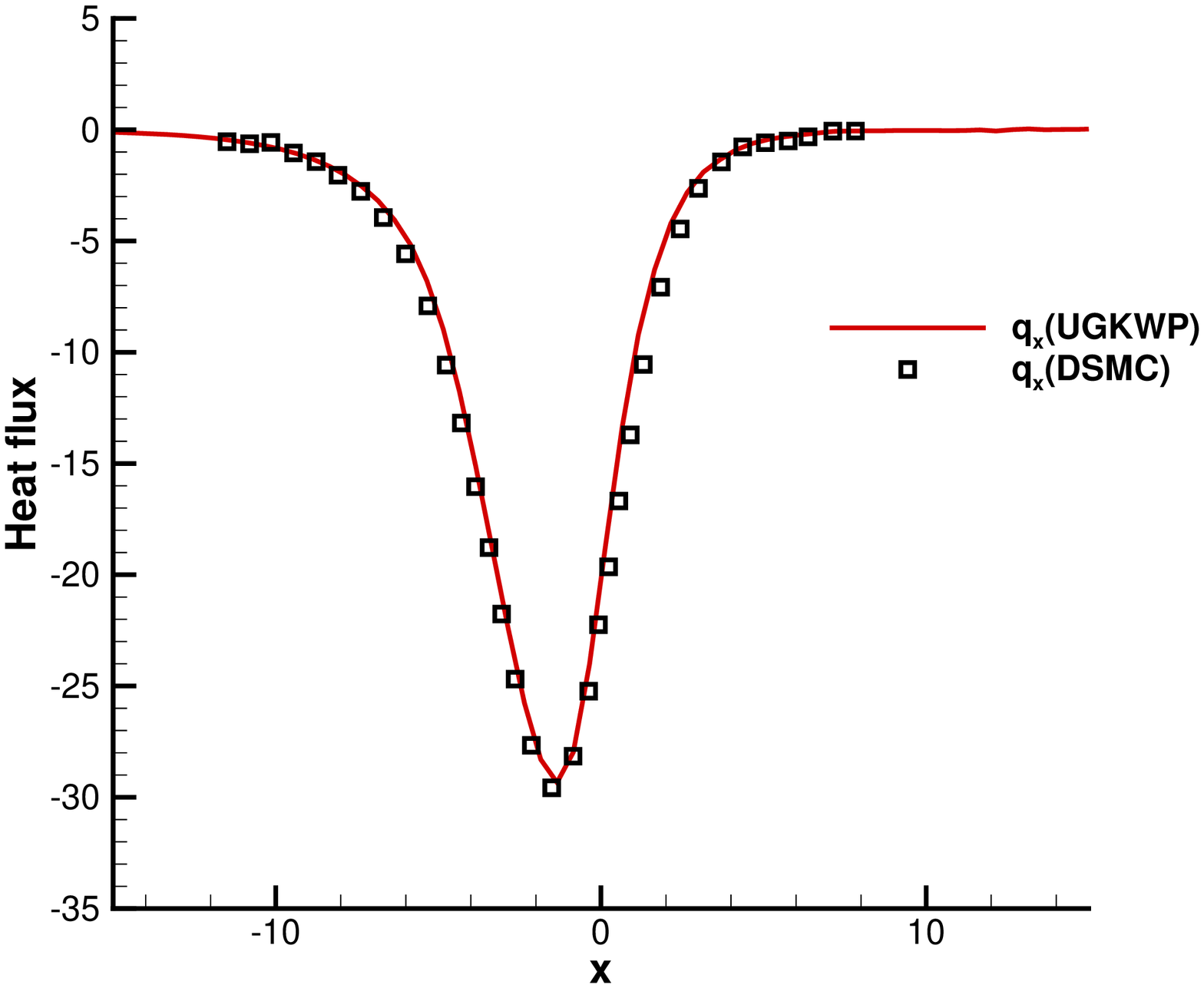}{b}
		\includegraphics[width=0.48\textwidth]{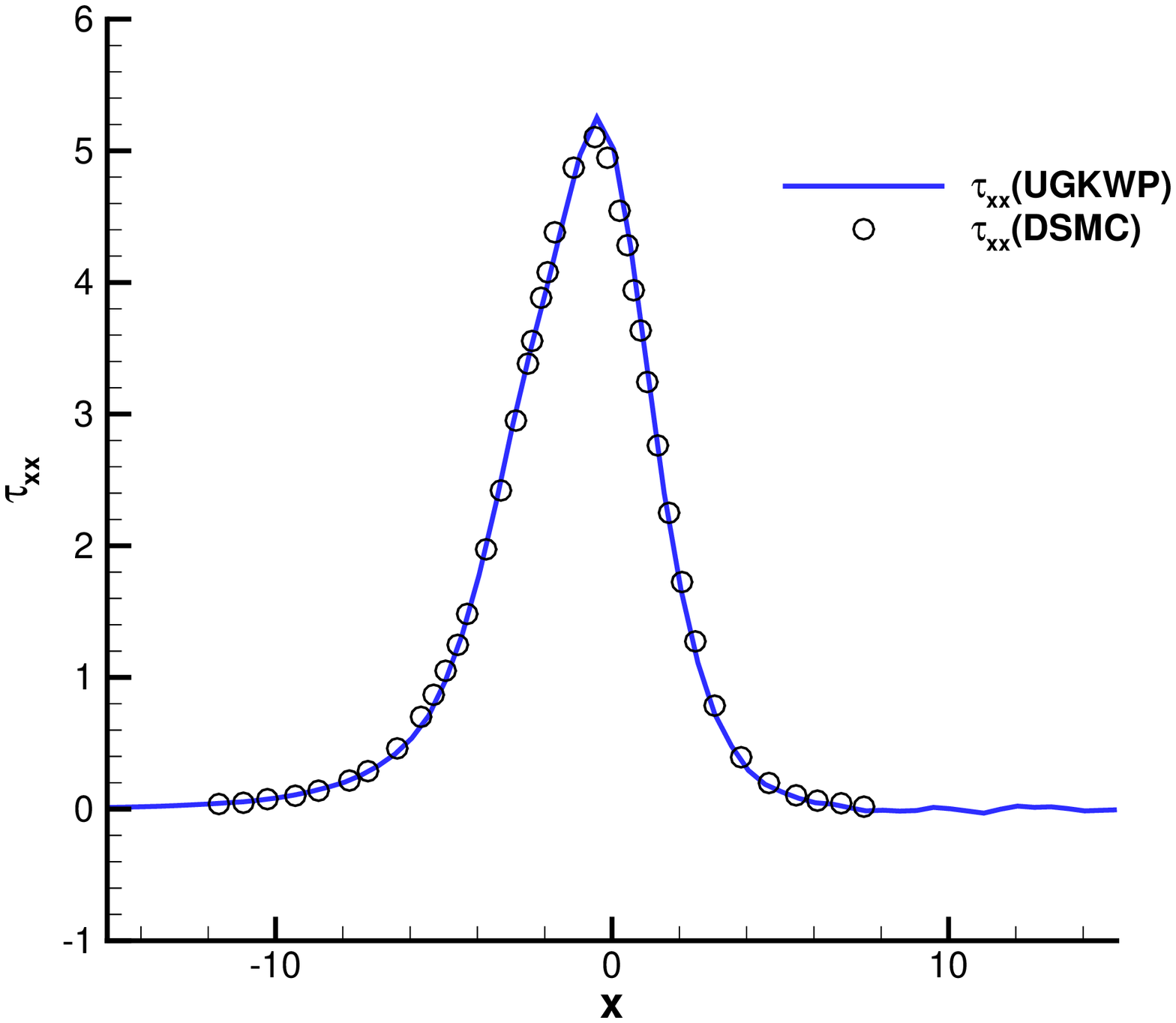}{b}
		\caption{$\mathrm{M} = 6.5$ argon shock structure calculated by the UGKWP and the DSMC. The x-coordinate is normalized by $\ell$.}
		\label{shockdsmc6x5}
	\end{figure}

\begin{figure}[htbp]
		\centering
		\includegraphics[width=0.48\textwidth]{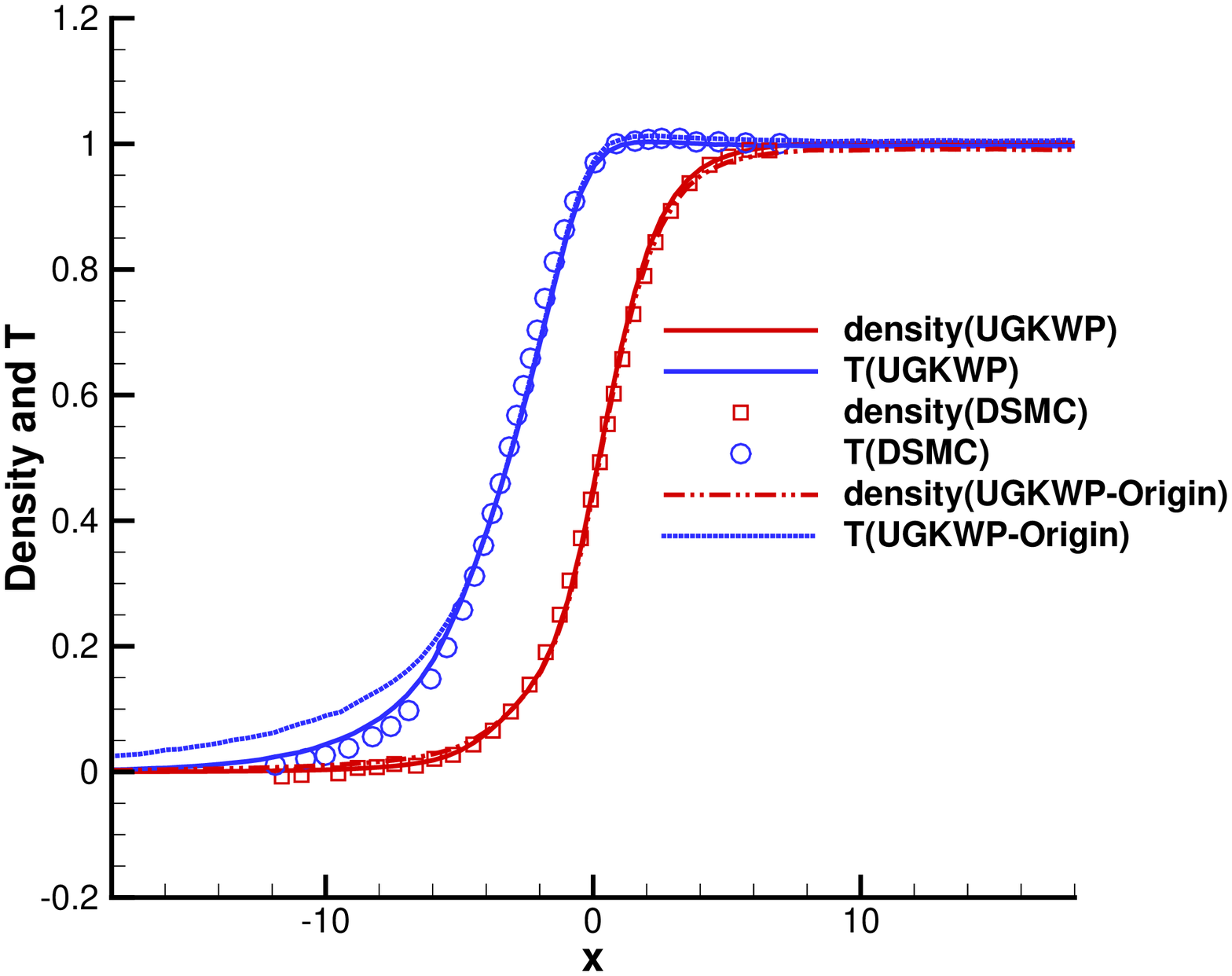}{a}
		\includegraphics[width=0.48\textwidth]{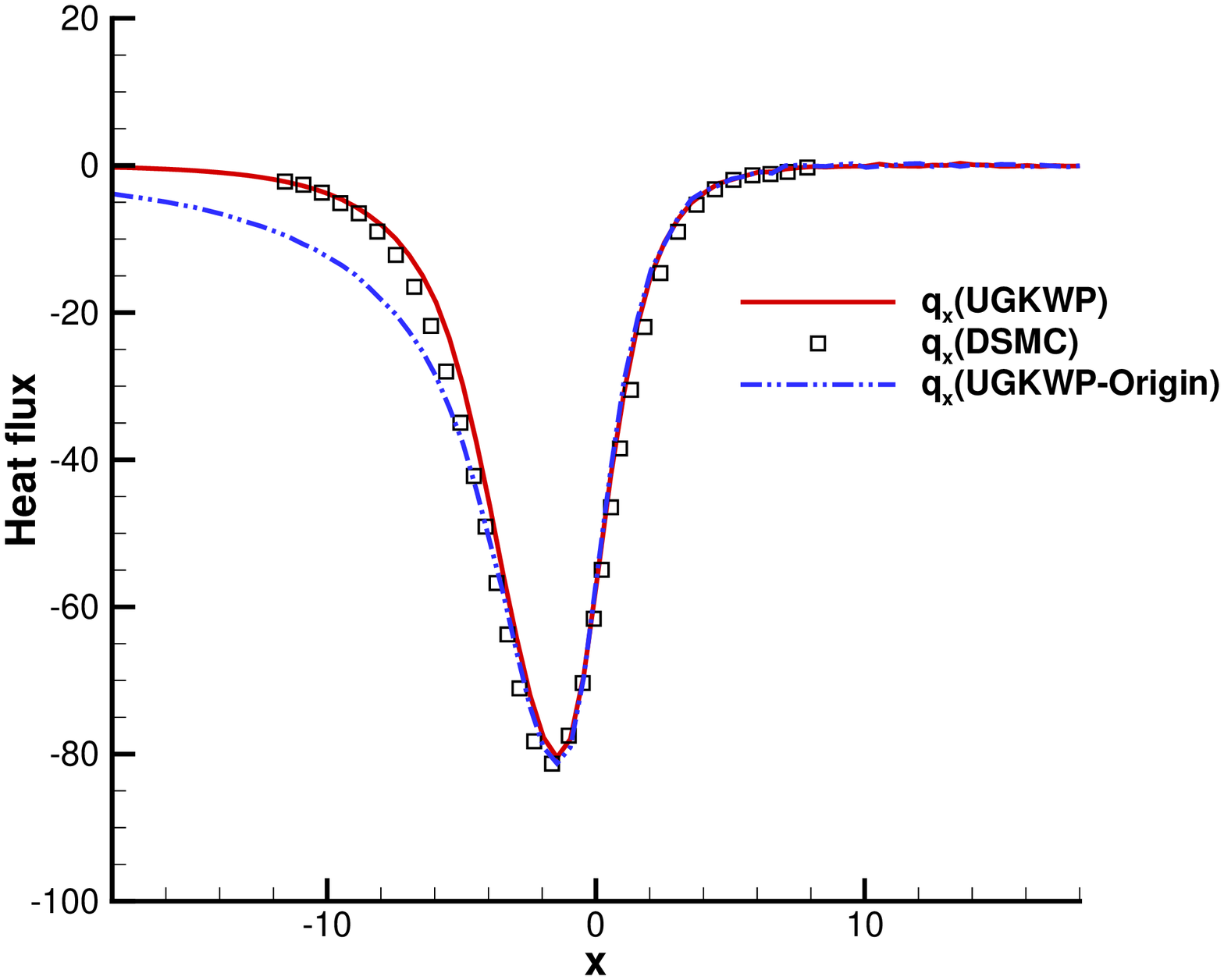}{b}
		\includegraphics[width=0.48\textwidth]{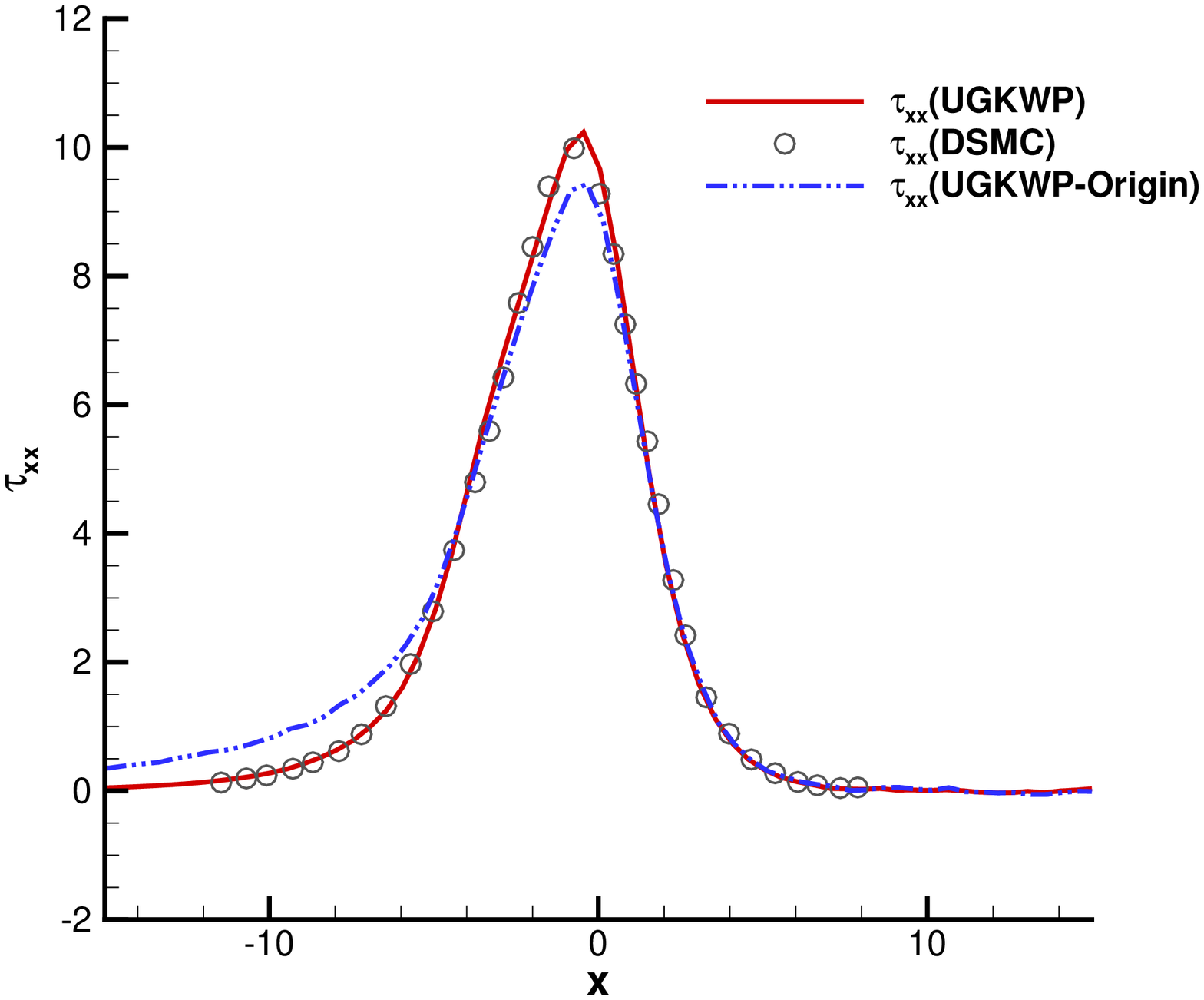}{c}
		\caption{$\mathrm{M} = 9$ argon shock structure calculated by the original and updated UGKWP and the DSMC. The x-coordinate is normalized by $\ell$.}
		\label{shockdsmc9}
	\end{figure}

	Fig. \ref{shockdsmc6x5} and \ref{shockdsmc9} show the argon shock structures for a viscosity coefficient of $\mu \sim T^{0.72}$ at Mach number $\mathrm{M} = 6.5$ and $9$, where the comparisons among original, updated UGKWP, and DSMC solutions are presented at $\mathrm{M} = 9$.
Good agreement has been obtained at both Mach numbers between UGKWP and the DSMC solutions.

\begin{figure}[htbp]
		\centering
		\includegraphics[width=0.48\textwidth]{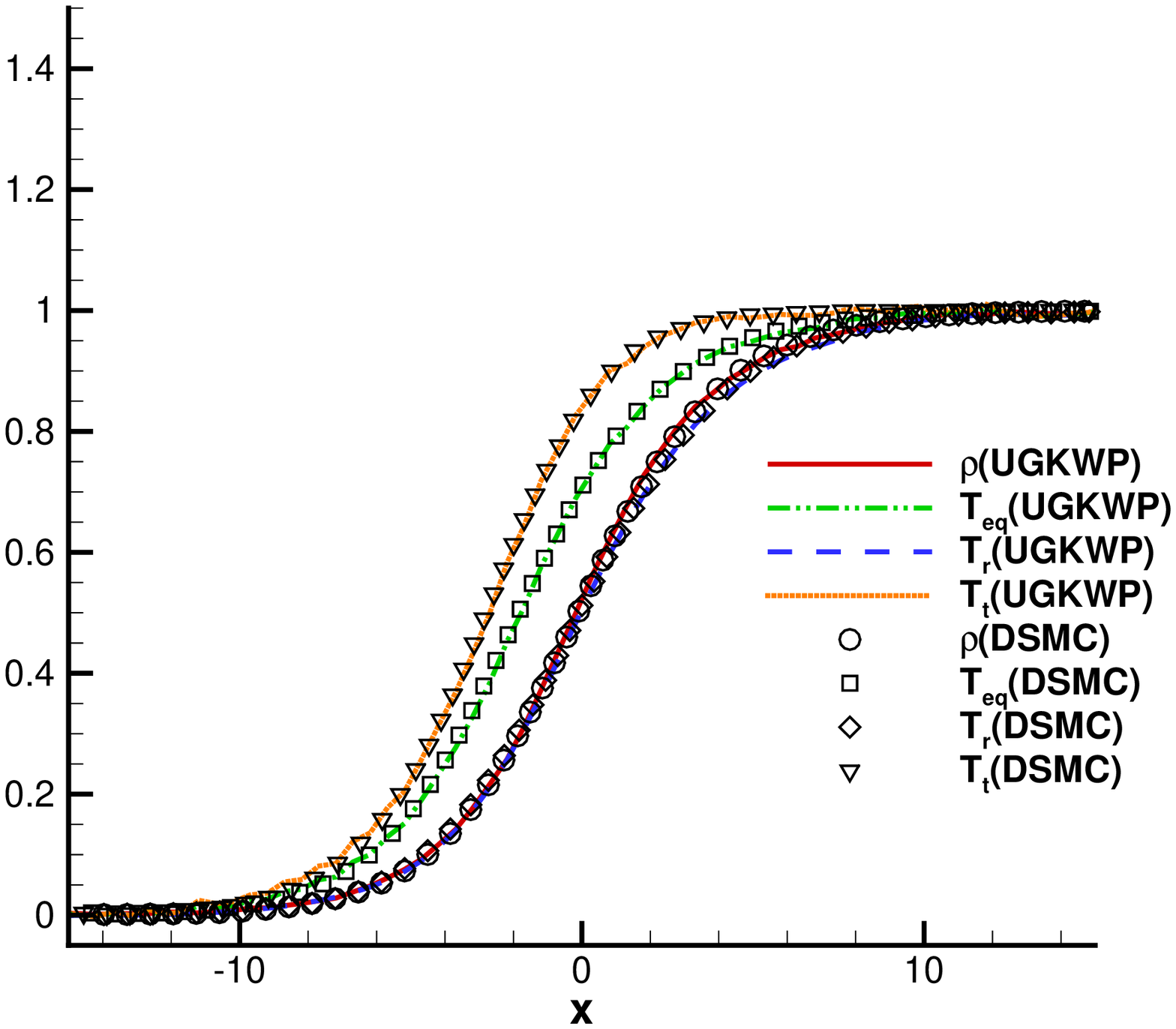}{a}
		\includegraphics[width=0.48\textwidth]{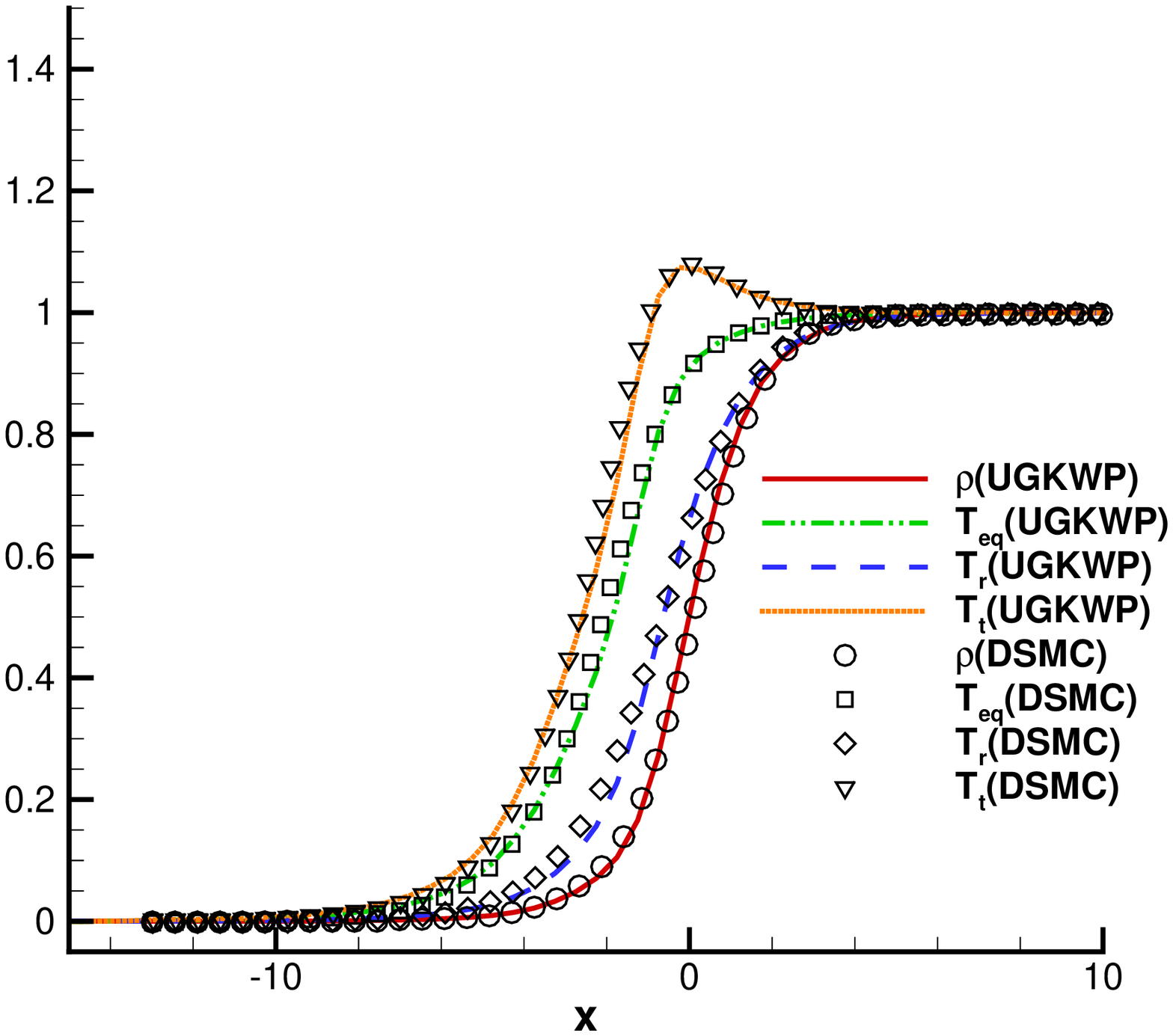}{b}
		\includegraphics[width=0.48\textwidth]{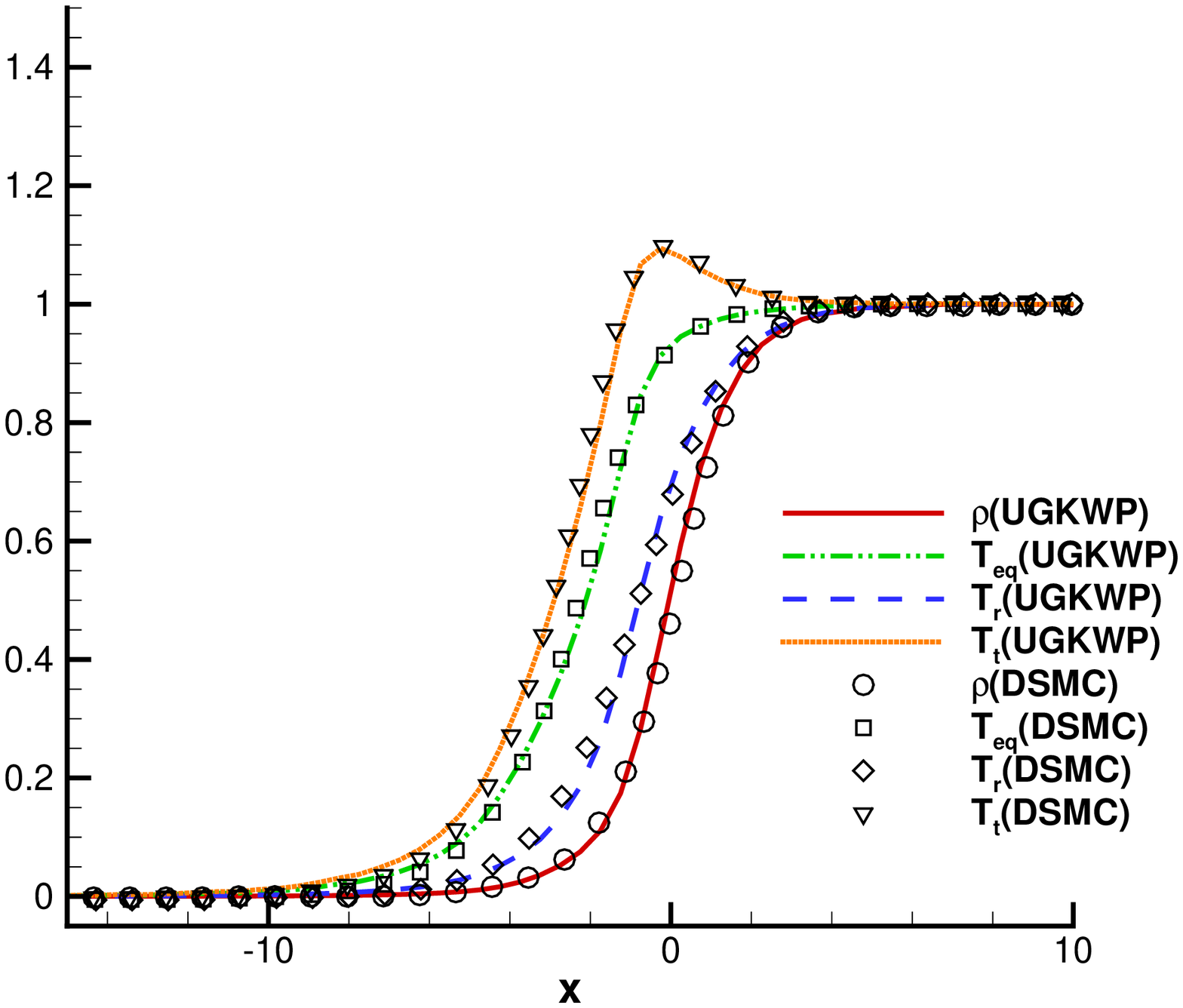}{c}
		\includegraphics[width=0.48\textwidth]{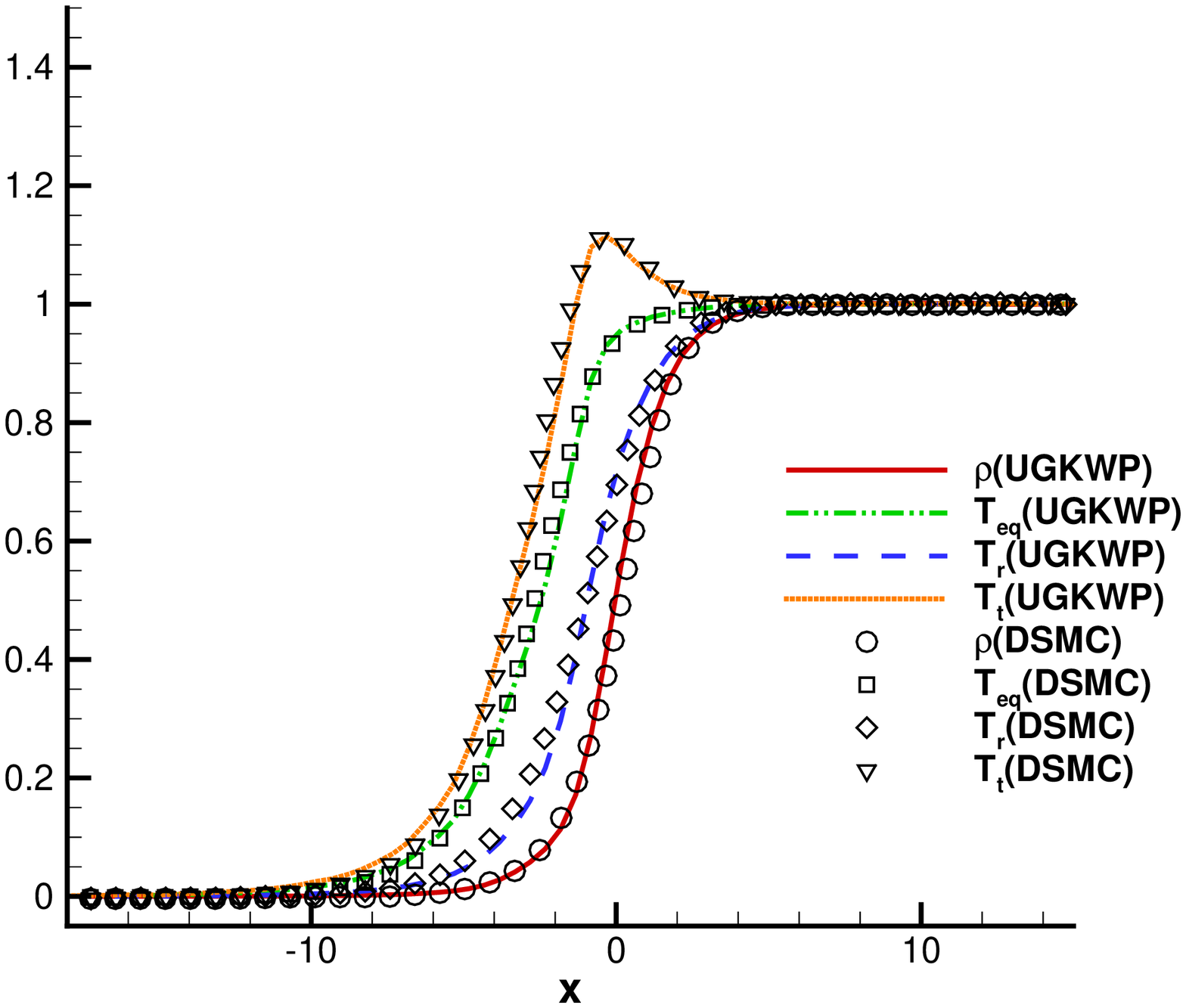}{d}
		\caption{Comparison of UGKWP and DSMC results of nitrogen shock wave at different Mach numbers for nitrogen gas. (a)$\mathrm{M} = 1.53$; (b)$\mathrm{M} = 4.0$; (c)$\mathrm{M} = 5.0$; (d)$\mathrm{M} = 7.0$. The x-coordinate is normalized by $\ell$.}
		\label{shockdsmc}
	\end{figure}

	In addition to monatomic gas, the results of shock structure in diatomic gas will be presented as well.
The diatomic gas UGKWP method is presented in \cite{xu-c}.
For diatomic gas, besides the translational relaxation, the rotational relaxation is included as well. The relaxation time between the rotational and translation energy exchange is determined by the rotational collision number $Z_{rot}$. For the nitrogen gas, the viscous coefficient follows Eq.\eqref{vhs-vs1} and Eq.\eqref{vhs-vs2} with the temperature dependent index $\omega=0.74$. In this calculation, the reference length is the upstream mean free path, and the computational domain is [-25,25] with 100 cells.The rotational collision number used in the UGKWP is $Z_{rot} = 2.4$. The normalized temperature and density comparison between UGKWP and DSMC  at $\mathrm{M} = 1.53, 4.0, 5.0 ,7.0$ are plotted in Fig. \ref{shockdsmc}. With  the modification of particle collision time, the UGKWP avoids the early temperature rising problem and presents good agreement with the DSMC result. In comparison with the Rykov model-based UGKS results \cite{liu2014unified}, significant improvement has been observed.

	In order to further validate the newly modified UGKWP method in the high speed rarefied flow regime,  the flow of argon gas passing through a circular cylinder at different Mach numbers is calculated. Two Mach numbers $\mathrm{M} = 10$ and $20$ with the Knudsen number $\mathrm{Kn} = 0.1$ will be tested. The Knudsen number is defined as the ratio of the mean free path over the cylinder radius. The radius of the cylinder is given by $R = 0.01m$. For $\mathrm{M} = 10$, the incoming argon gas has a velocity $U_{\infty} = 3077.587m/s$, the temperature $T_{\infty} = 273K$, molecular number density $n_{\infty} = 1.2944\times 10^{21}/m^3$, and the reference viscosity $\mu_{\infty} = 2.117\times 10^{-5} Ns/m^2$. The viscosity is  calculated by Eq.\eqref{vhs-vs1} with $\omega = 0.81$. The cylinder has constant surface temperature $T_w = 273K$, and diffusive boundary condition is adopted here. For $\mathrm{M} = 20$, the only change is the incoming velocity with a value of $U_\infty = 6155.174m/s$.
	
\begin{figure}[htbp]
		\centering
		\includegraphics[width=0.48\textwidth]{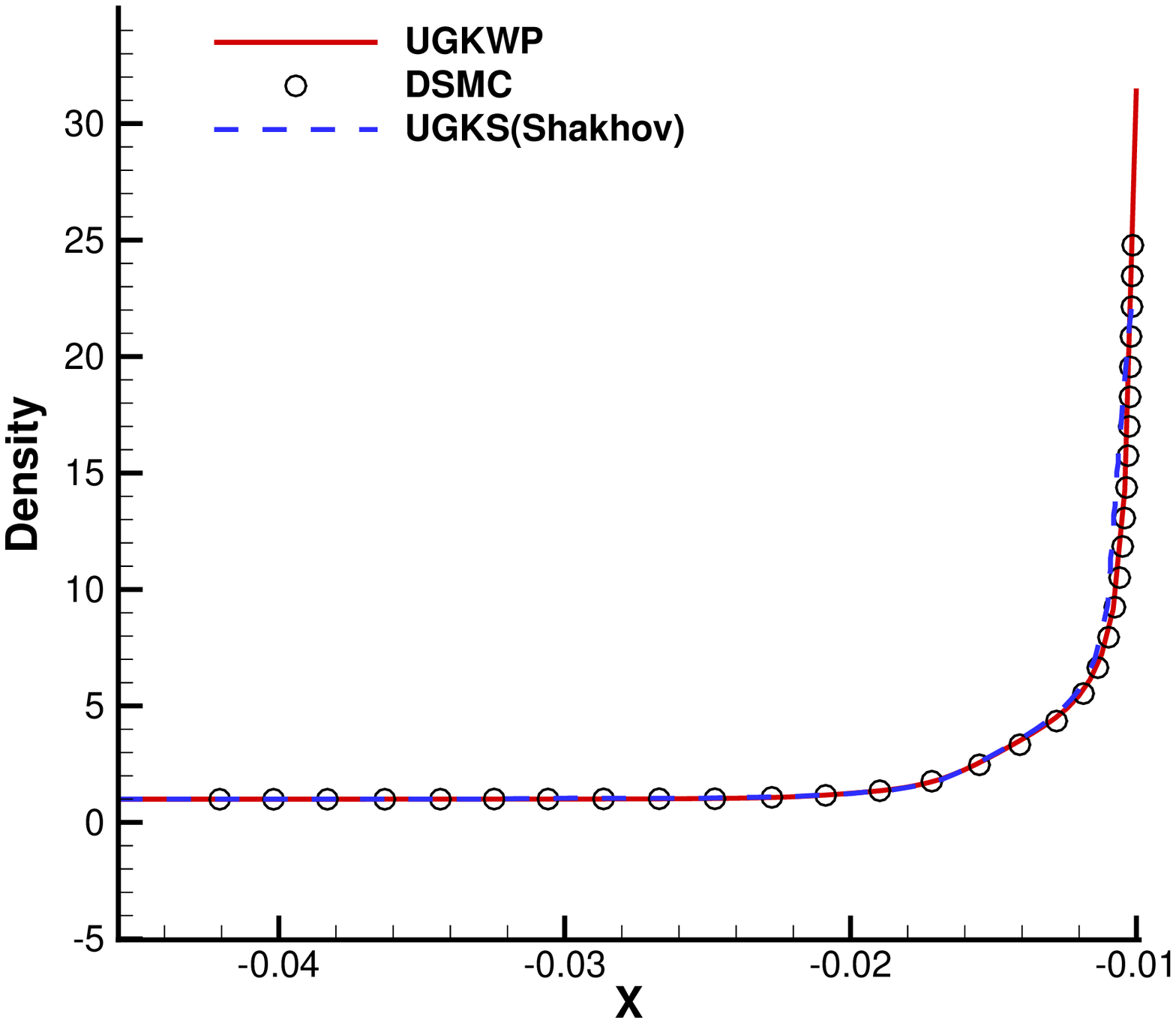}{a}
		\includegraphics[width=0.48\textwidth]{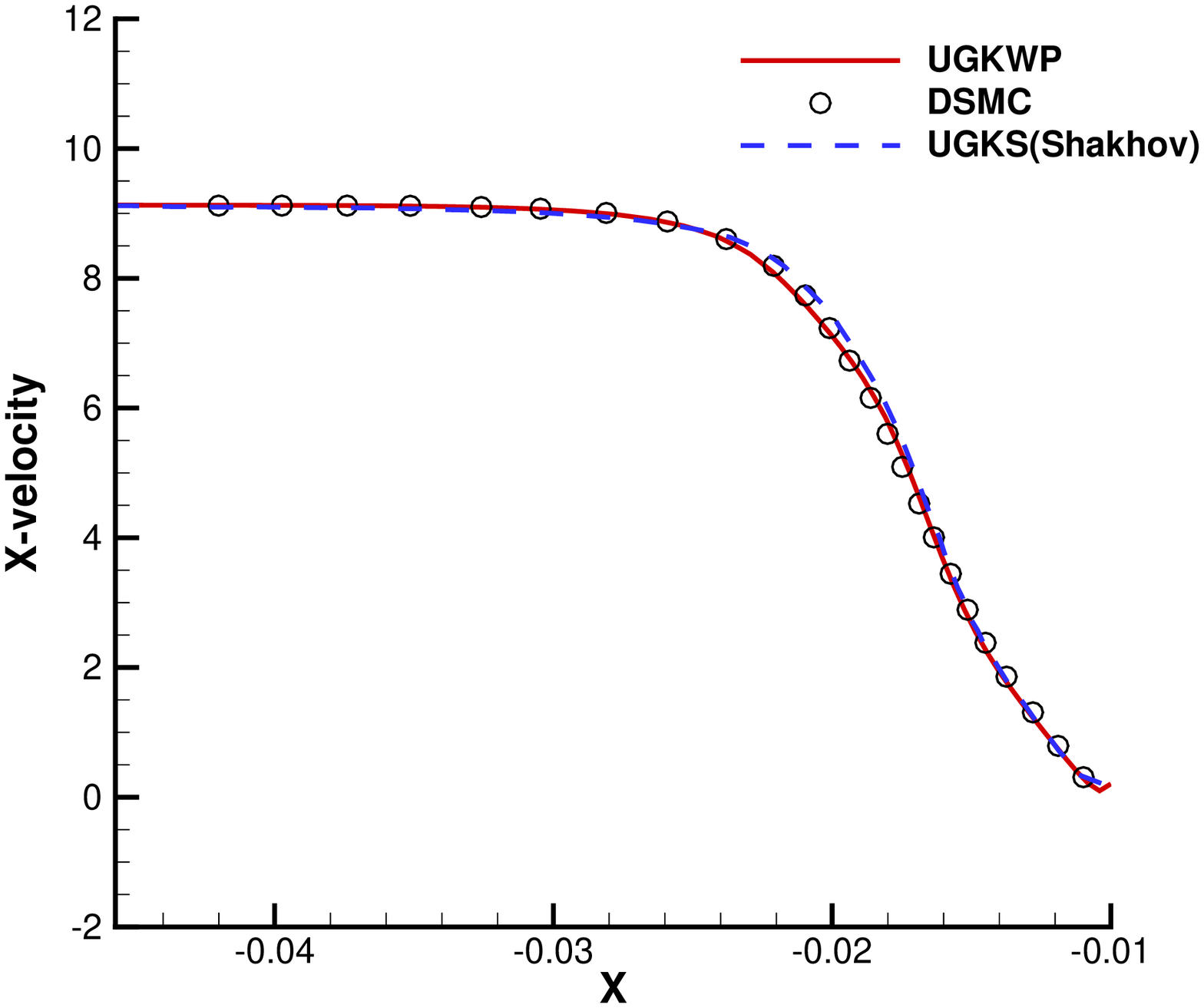}{b}
		\includegraphics[width=0.48\textwidth]{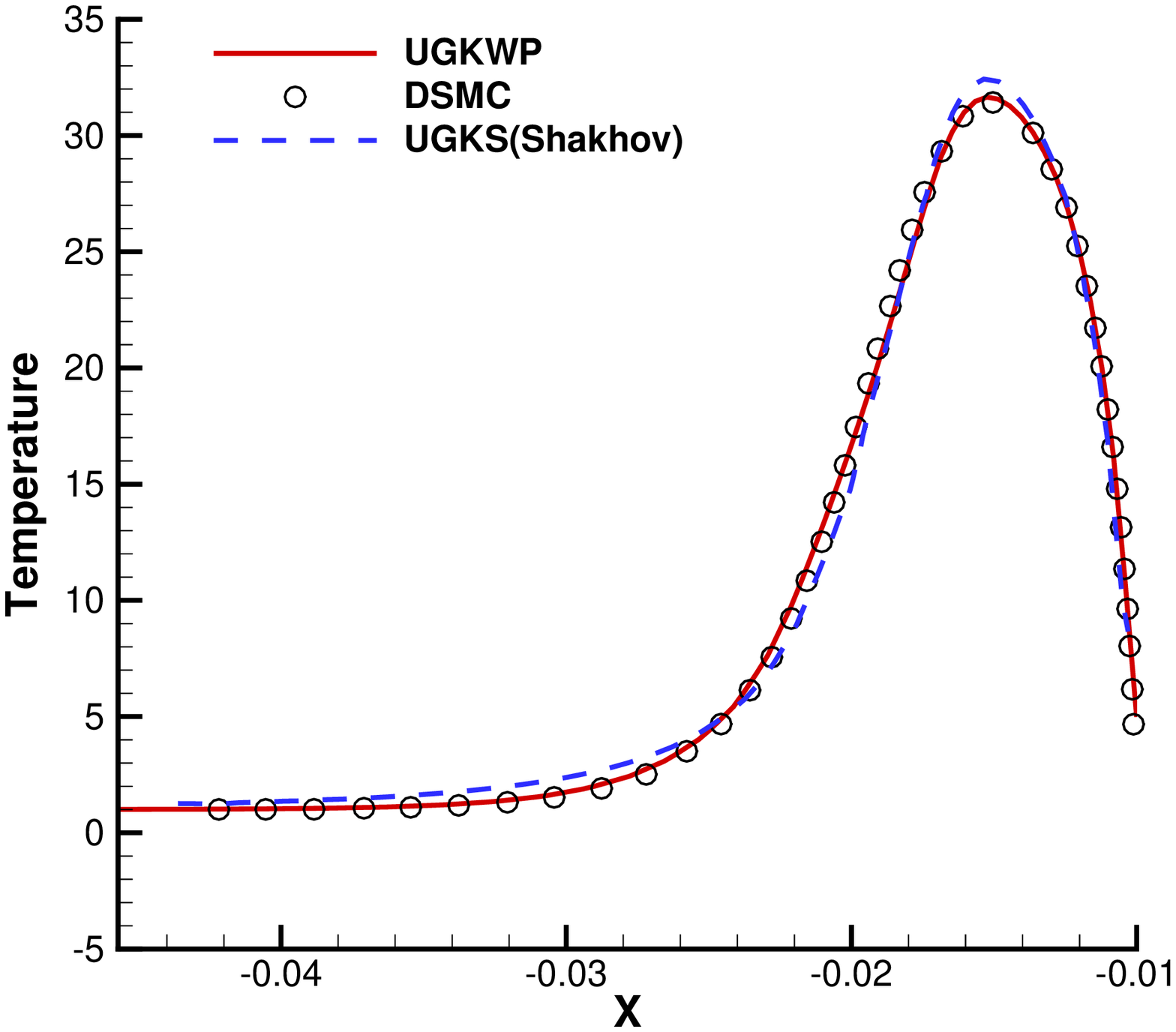}{c}
		\caption{Flow distributions for argon gas along the central symmetric line in front of the stagnation point at $\mathrm{M = 10}$ and $\mathrm{Kn} = 0.1$.}
		\label{Ma10cylinder}
	\end{figure}

\begin{figure}[htbp]
	\centering
	\includegraphics[width=0.48\textwidth]{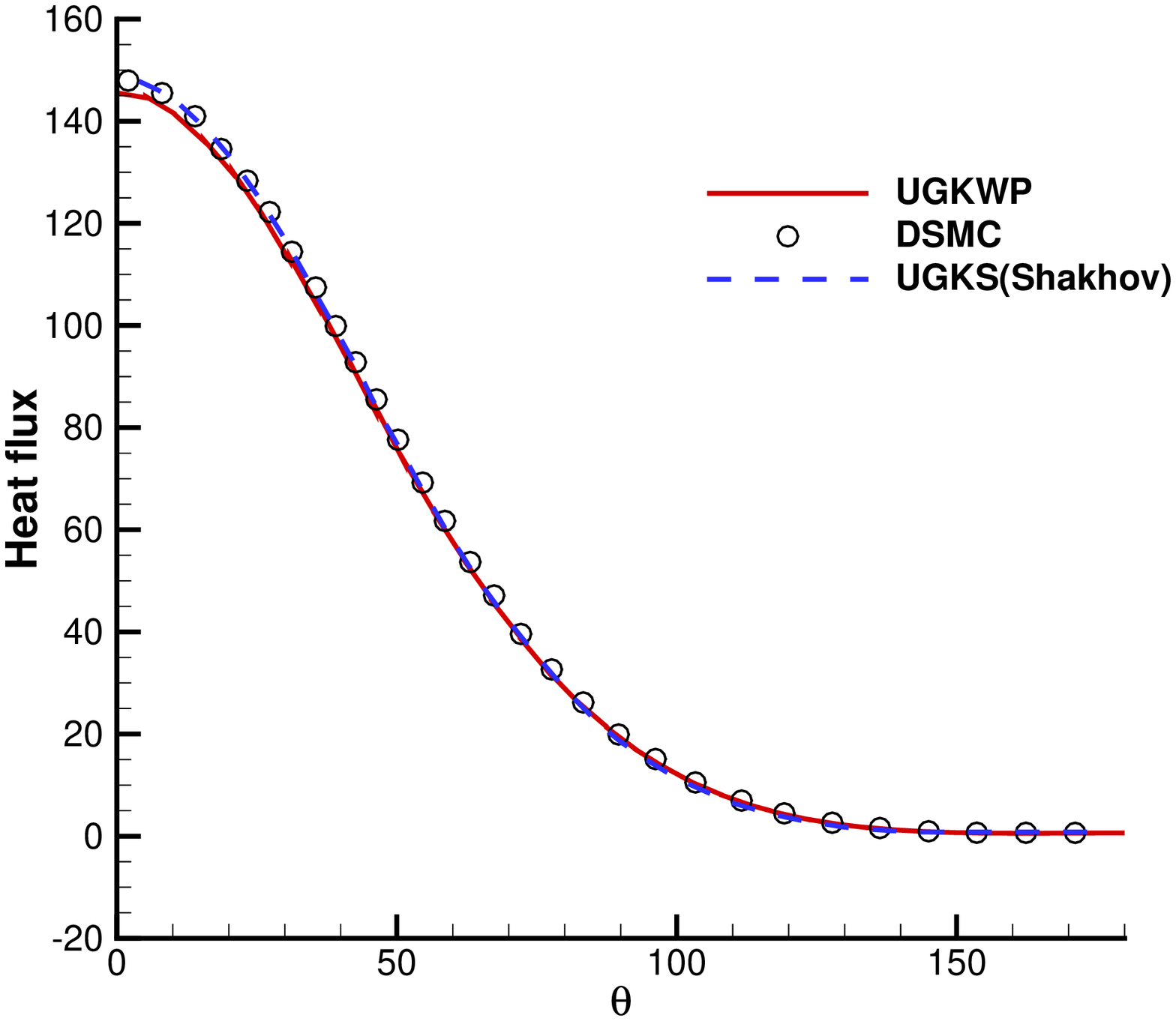}{a}
	\includegraphics[width=0.48\textwidth]{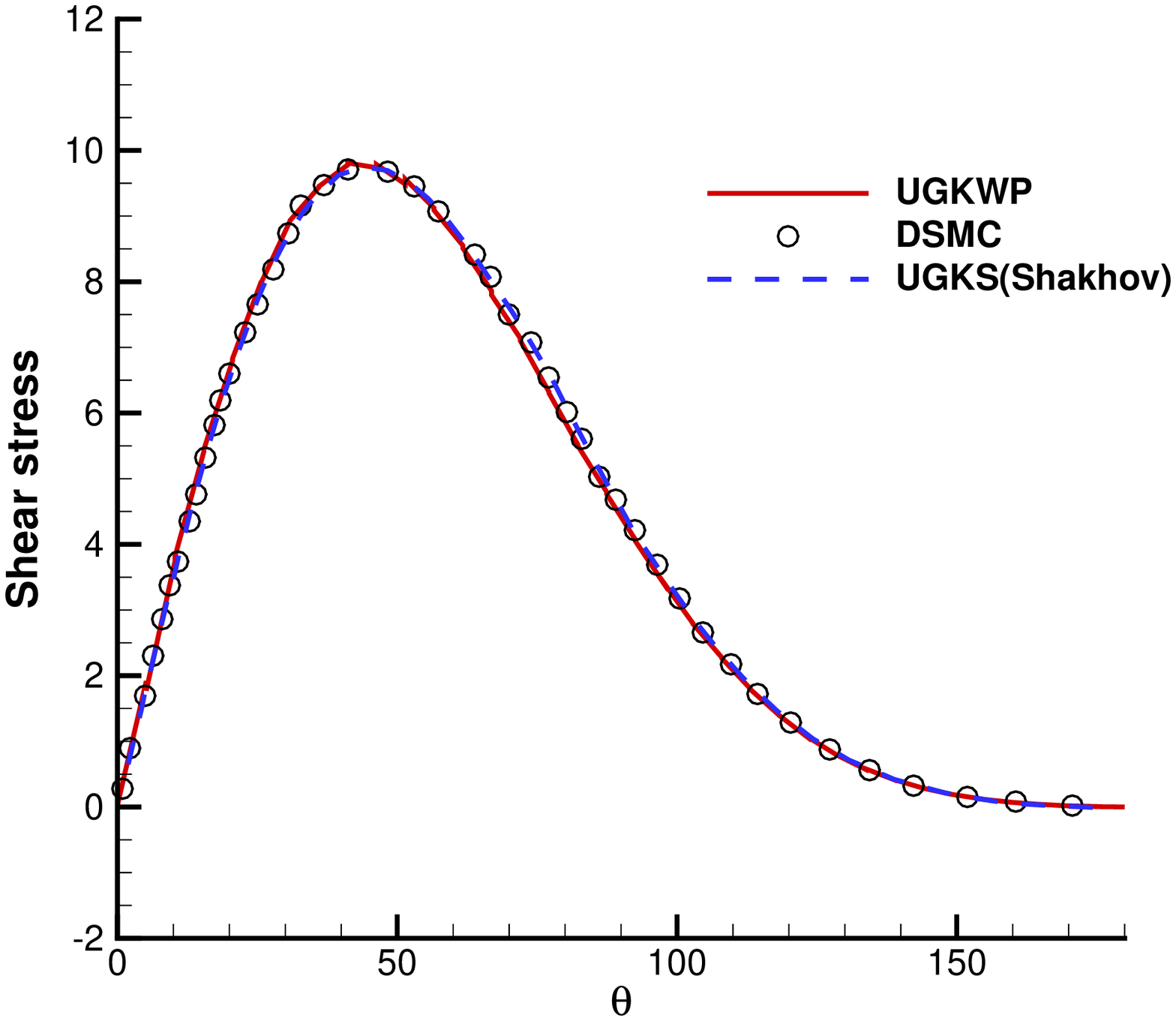}{b}
	\includegraphics[width=0.48\textwidth]{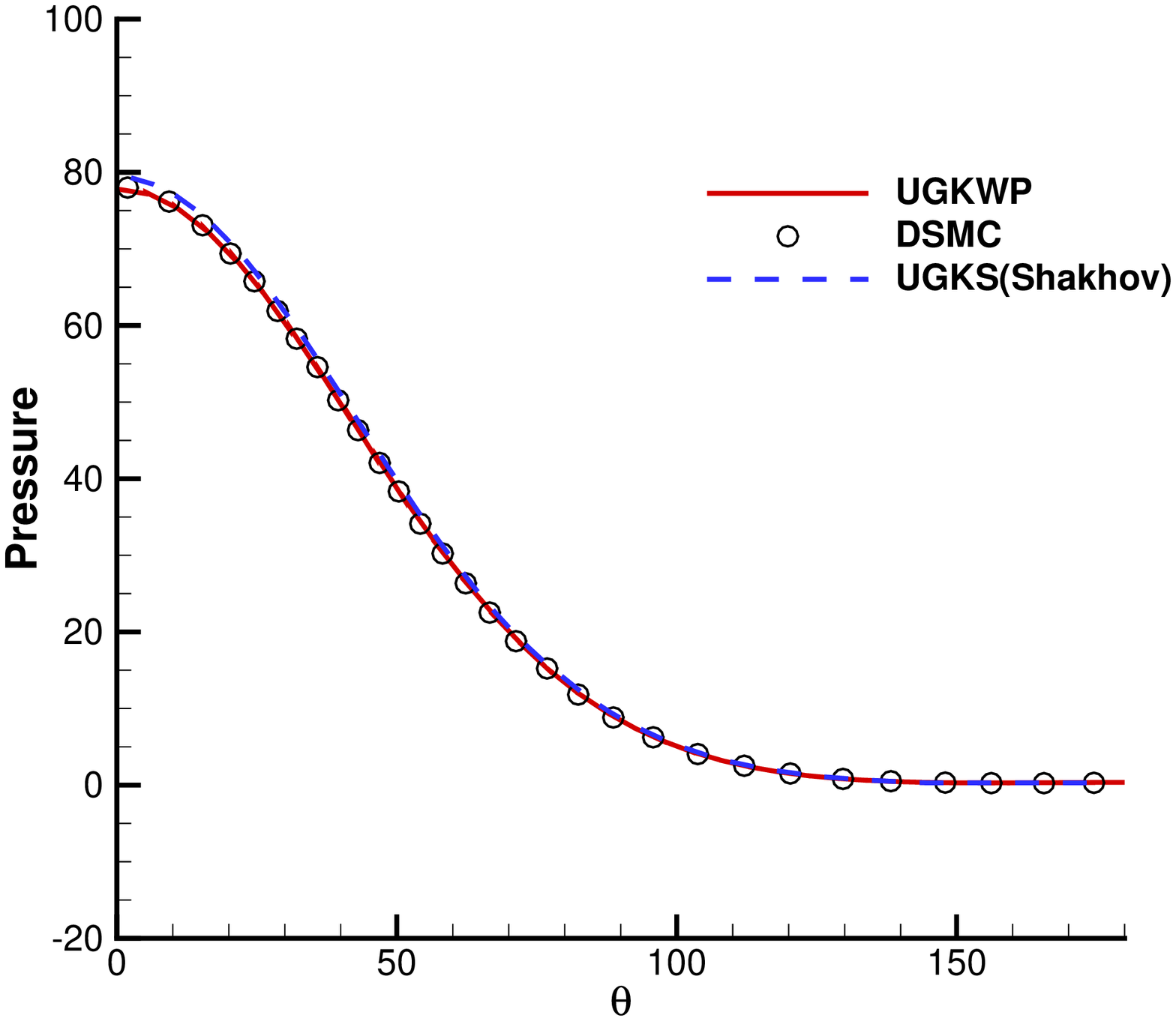}{c}
	\caption{Surface quantities along the surface of the cylinder for argon gas at $\mathrm{M = 10}$ and $\mathrm{Kn} = 0.1$. }
	\label{Ma10cylinderSurface}
\end{figure}

\begin{figure}[htbp]
	\centering
	\includegraphics[width=0.48\textwidth]{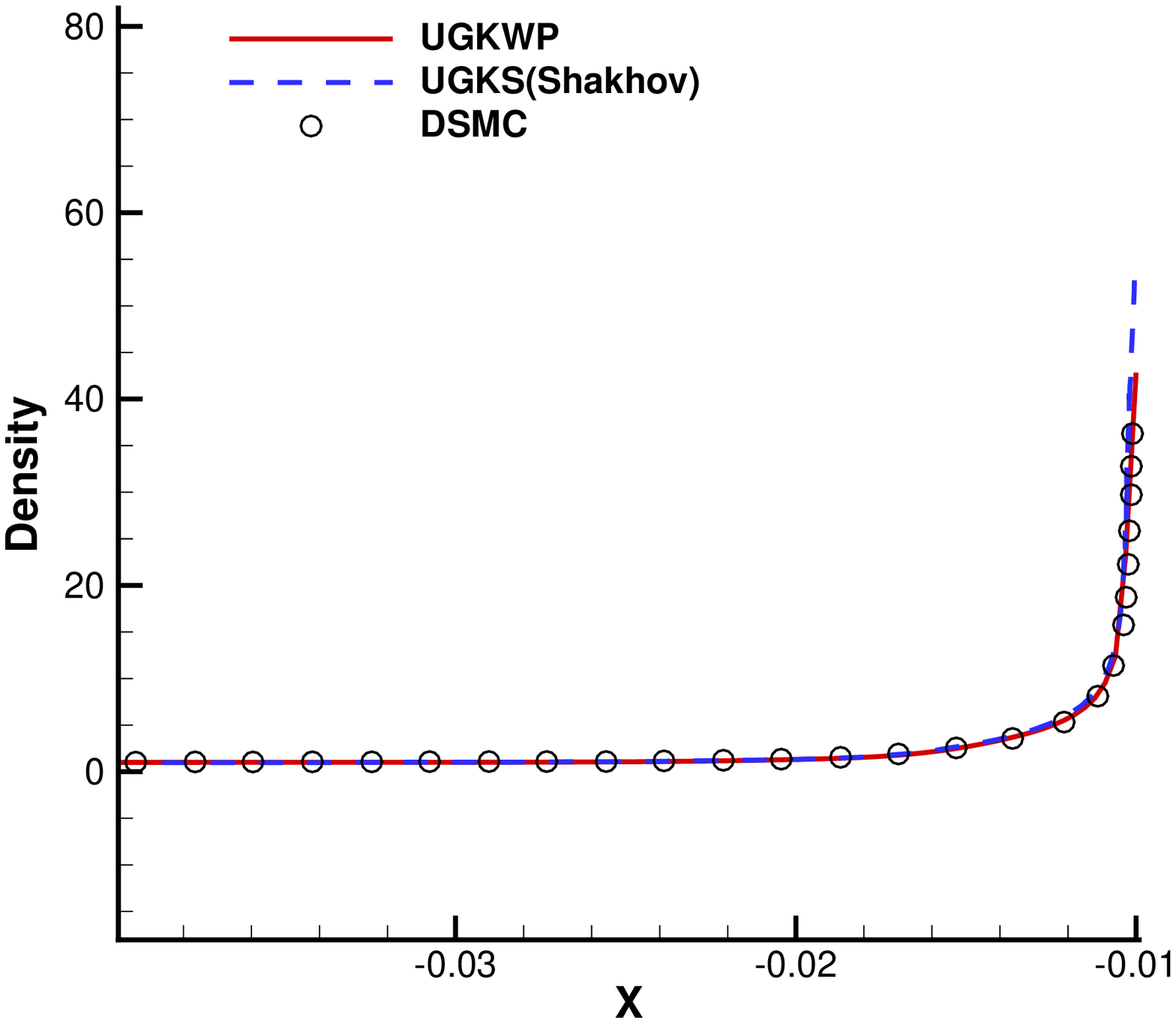}{a}
	\includegraphics[width=0.48\textwidth]{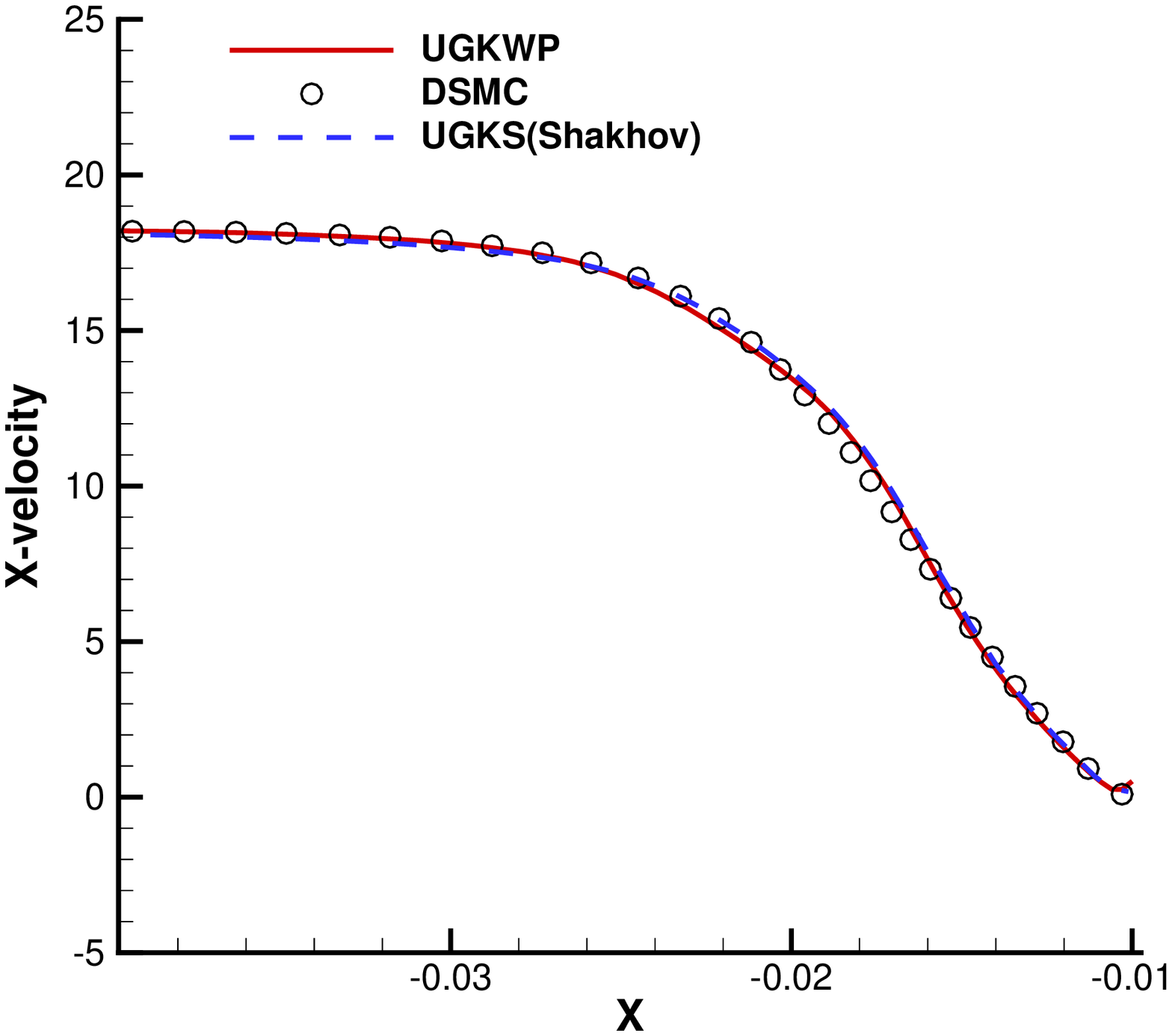}{b}
	\includegraphics[width=0.48\textwidth]{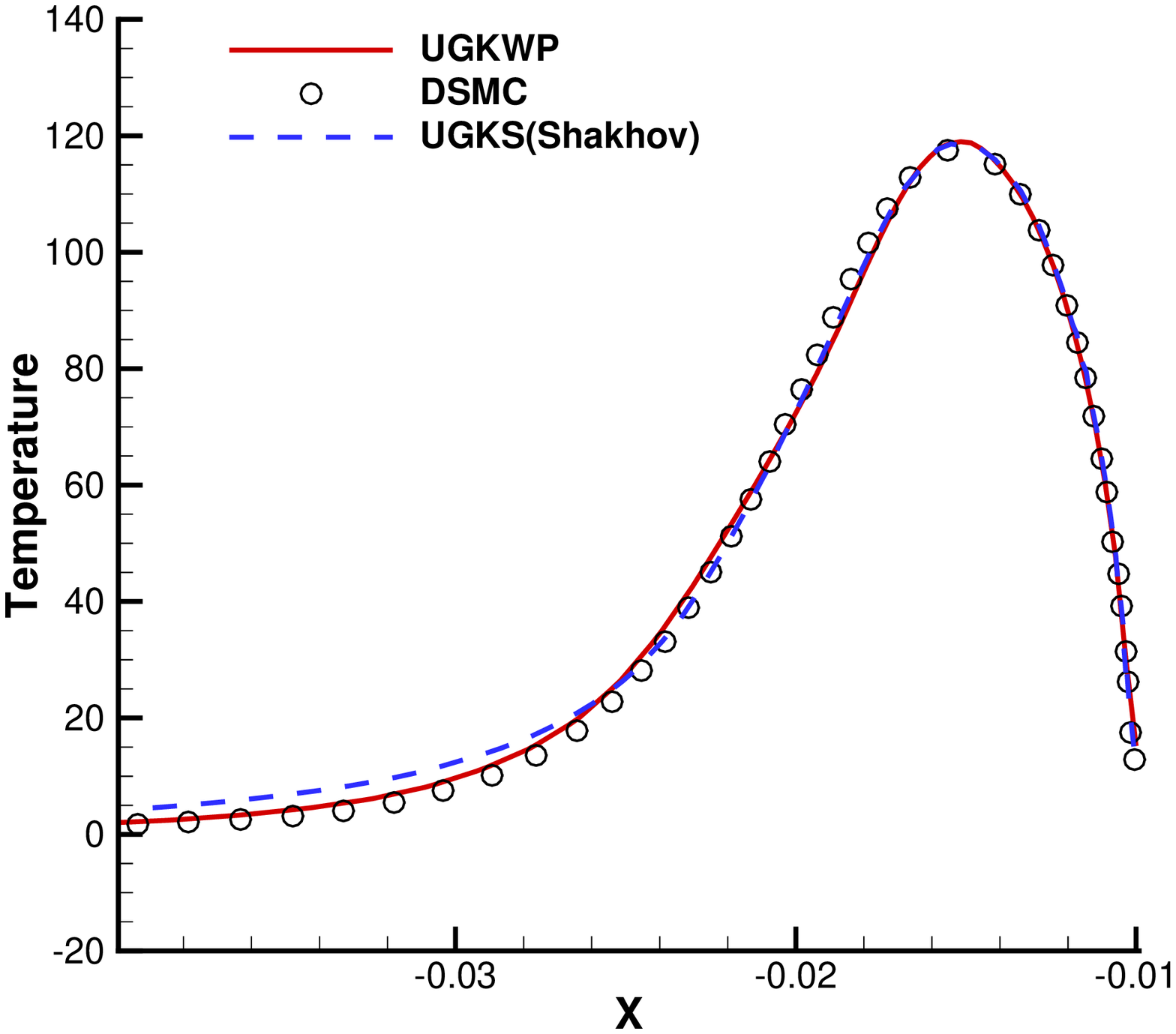}{c}
	\caption{Flow distributions for argon gas along the central symmetric line in front of the stagnation point at $\mathrm{M = 20}$ and $\mathrm{Kn} = 0.1$.}
	\label{Ma20cylinder}
\end{figure}

\begin{figure}[htbp]
	\centering
	\includegraphics[width=0.48\textwidth]{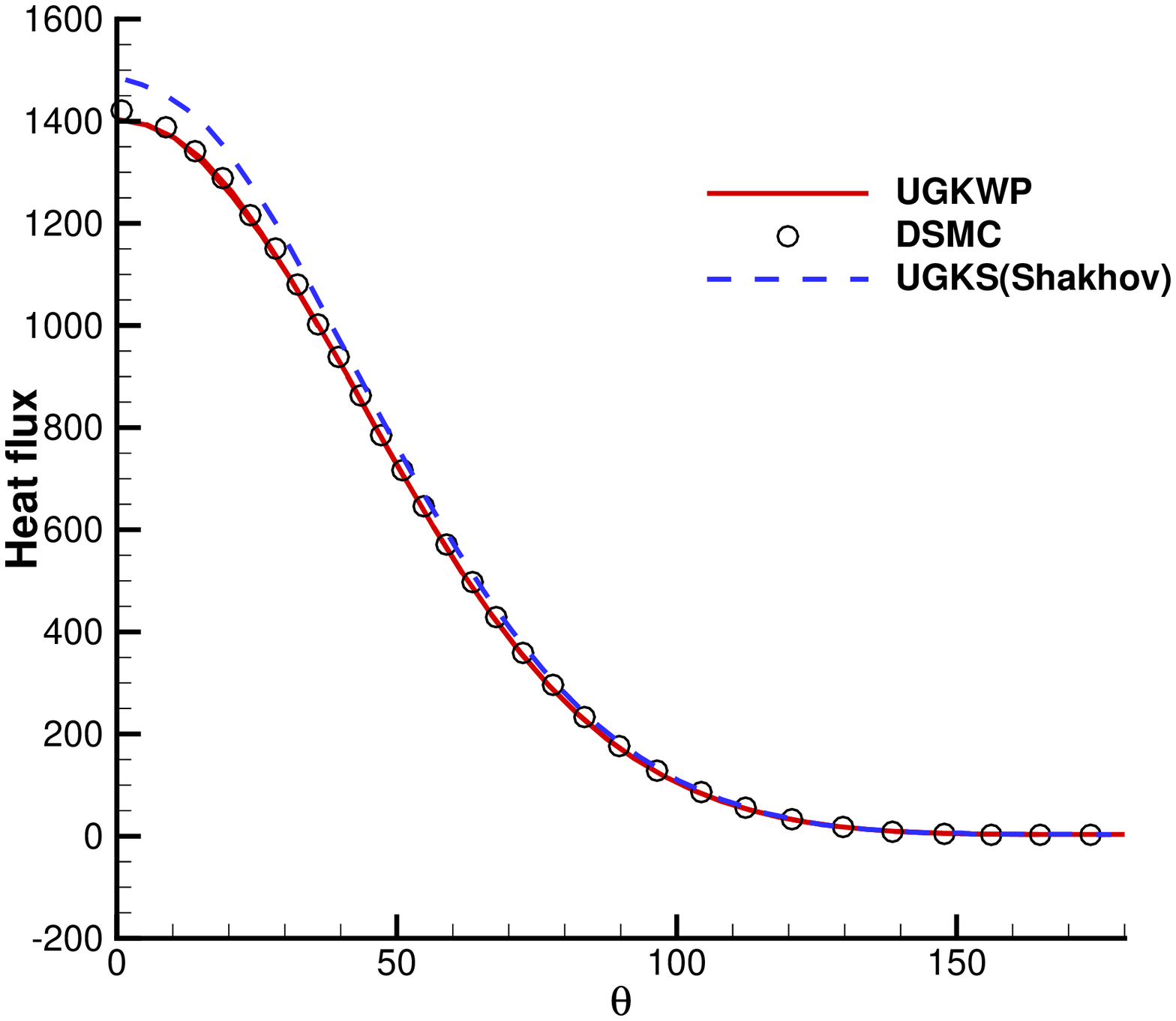}{a}
	\includegraphics[width=0.48\textwidth]{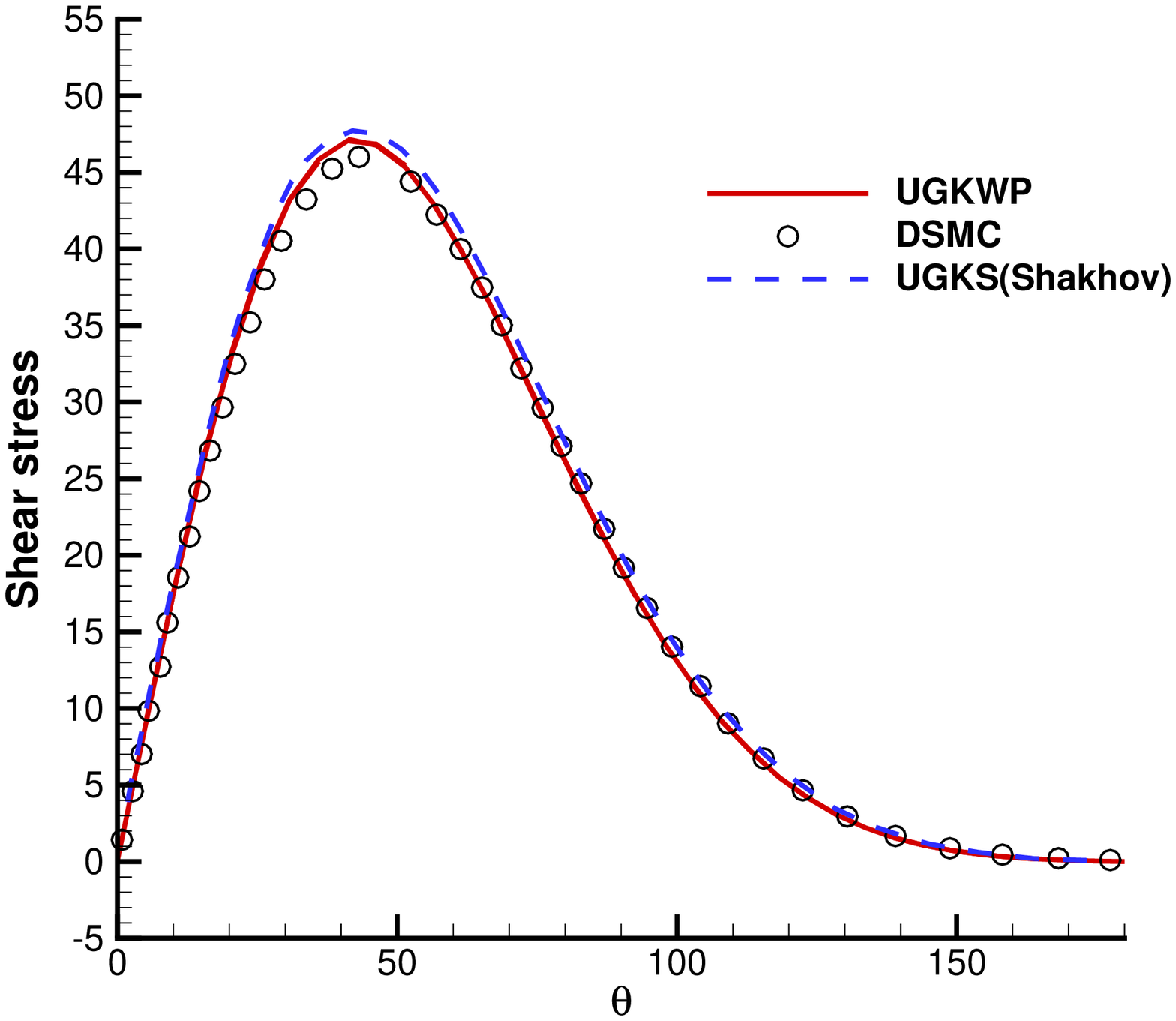}{b}
	\includegraphics[width=0.48\textwidth]{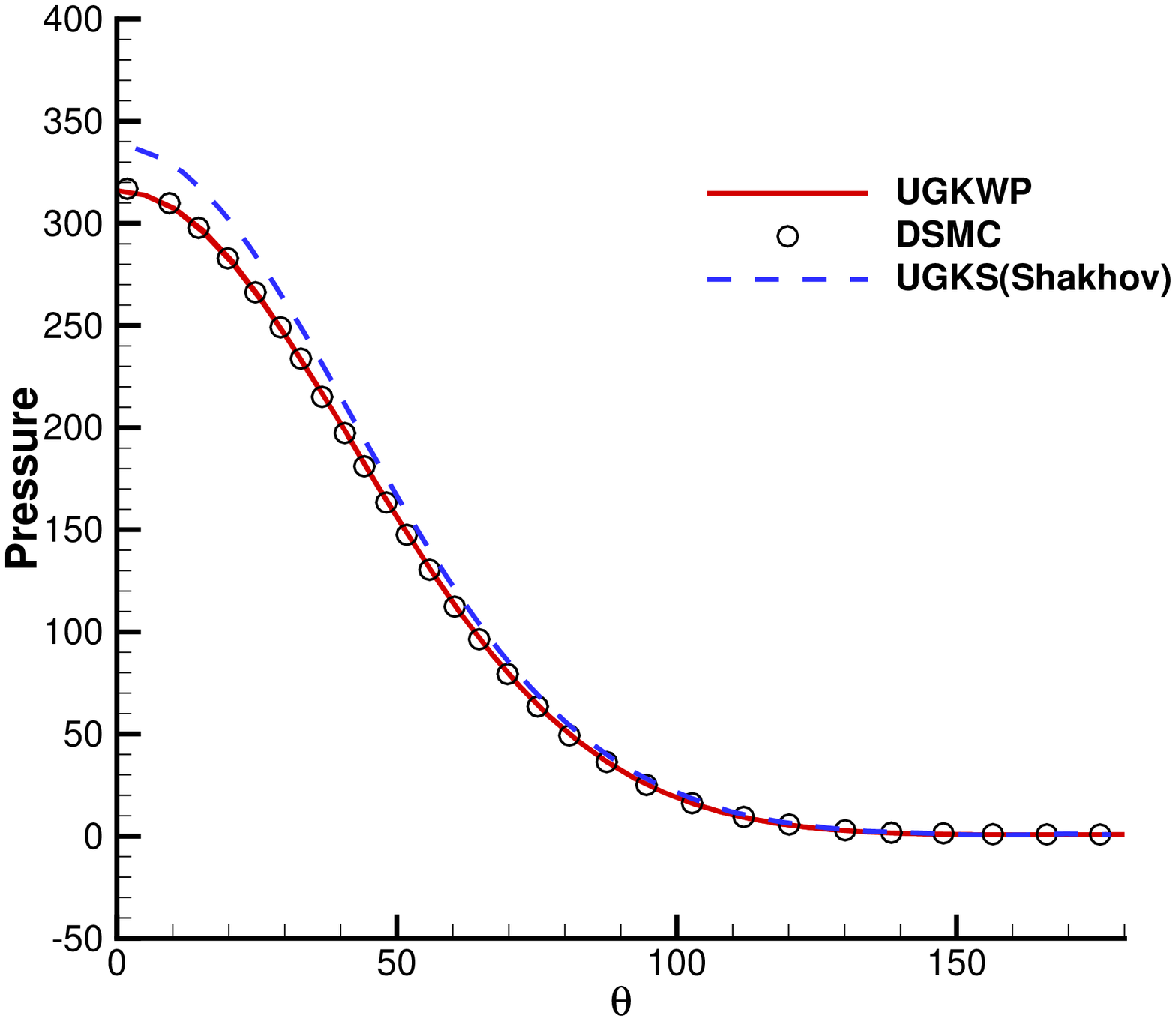}{c}
	\caption{Surface quantities along the surface of the cylinder for argon gas at $\mathrm{M = 20}$ and $\mathrm{Kn} = 0.1$. }
	\label{Ma20cylinderSurface}
\end{figure}

	For both cases, the simulation results from UGKWP are compared the solutions of the DSMC and the UGKS-Shakhov \cite{yu}. 	
For $\mathrm{M} = 10$ case, Fig. \ref{Ma10cylinder} shows the density, x-velocity, and temperature along the central symmetric line in front of the stagnation point, where both UGKWP and UGKS-Shakhov solutions agree with the DSMC solutions, except that the temperature in UGKS-Shakhov solutions rises  a little earlier. The comparison of the heat flux, shear stress, and pressure along the surface of the cylinder are shown in Fig. \ref{Ma10cylinderSurface}, which have good agreement with DSMC solution.
		For $\mathrm{M} = 20$ case, the results  are plotted in Fig. \ref{Ma20cylinder} and \ref{Ma20cylinderSurface}. As shown in Fig. \ref{Ma20cylinder}, the differences in temperature distributions from the UGKS-Shakhov and the DSMC are much more obvious than that in the $M=10$ case, where the results from the modified UGKWP have good agreement with the DSMC solution, especially for the pressure and heat flux distributions along
 the surface of the cylinder.

	\section{Conclusion}
	
	In this paper,  a newly modified UGKWP method has been proposed for modeling and computation of non-equilibrium flow.
The main idea is to adjust particle collision time according particle velocity, which is more consistent with the kinetic theory than the
single relaxation kinetic model. The modeling in UGKWP is beyond the traditional BGK-type kinetic models.
	With the implementation of particle velocity-dependent collision time, the non-equilibrium solution can be captured accurately,
    such as the shock structure calculations.
    Based on the simulation results, no obvious discrepancy between the UGKWP and DSMC results can be observed.
    Intrinsically, the DSMC collision rate
	is determined by the relative particle velocity, which has the similarity with the modification of collision time according to particle velocity here.
	Since the dynamic effect from most particles around the average velocity has been taken into account in the kinetic relaxation model,
	only the collision time for these particles with extremely high speed needs to be modified.
Under such a modeling, the UGKWP method is still fully conservative. The only modification is the distance travelled by the very high speed particles.
	The additional work introduced in the modification of collision time doesn't increase any computational cost of UGKWP.
		In the continuum flow regime, the UGKWP will automatically recover the gas-kinetic scheme for the Navier-Stokes solutions with the evolution of macroscopic flow variables alone, while the flux from the gas distribution has an analytical formulation without using particles.
For the flow simulation with the co-existing  multiple flow regimes,
the UGKWP method can achieve high efficiency and present very accurate solution in each flow regime.

	\section*{Acknowledgments}
	The current research is supported by National Numerical Windtunnel project and  National Science Foundation of China 11772281, 91852114.

\newpage





\section*{References}
	
	\begin{thebibliography}{99}
		
		
		
		\bibitem{aristov}
		{\sc V.V. Aristov}, {\em Direct Methods for Solving the Boltzmann
			Equation and Study of Nonequilibrium Flows}, Kluwer Academic
		Publishers (2001).
		
%		\bibitem{alsmeyer} {\sc H. Alsmeyer},
%		{\em Density profiles in argon and nitrogen shock waves measured by
%			the absorption of an eletron beam}, J. Fluid Mech., {\bf 74} (1976),
%		pp. 497.
		
		
		
		
		\bibitem{bgk} {\sc P.L. Bhatnagar, E.P. Gross, and M. Krook},
		{\em A Model for Collision Processes in Gases I: Small Amplitude
			Processes in Charged and Neutral One-Component Systems}, Phys. Rev.,
		{\bf 94} (1954), pp. 511-525.
		
		
		\bibitem{bird70} {\sc G.A. Bird},
		{\em Aspects of the structure of strong shock waves}, Phys. Fluids,
		{\bf 13} (1970), pp. 1172-1177.
		
		
		\bibitem{bird} {\sc G.A. Bird},
		{\em Molecular Gas Dynamics and the Direct Simulation of Gas Flows},
		Oxford Science Publications (1994).
		
		
		
		\bibitem{chapman} {\sc S. Chapman and T.G. Cowling},
		{\em The Mathematical Theory of Non-uniform Gases}, Cambridge
		University Press (1990).
		
		\bibitem{chu} {\sc C.K. Chu},
		{\em Kinetic-Theoretic Description of the Formation of a Shock
			Wave}, {\sl Phys. Fluids} {\bf 8}, 12 (1965).
		
		
		
%		\bibitem{erwin} {\sc D.A. Erwin, G.C. Pham-Van-Diep, and E.P. Muntz},
%		{\em Nonequilibrium gas flows. I: a detailed validation of Monte
%			Carlo simulation for monatomic gases}, {\sl Phys. Fluids A} {\bf
%			3}(4)(1991), pp. 697-705.
		
		\bibitem{holway} {\sc L. H. Holway Jr},
		{\em New statistical models for kinetic theory: methods of construction}, {\sl Phys. Fluids } {\bf
			9}(9)(1966), pp. 1658-1673.

         \bibitem{hu} {\sc Z.C. Hu and Z.N. Cai},
		{\em Burnett spectral method for high-speed rarefied gas flows}, {\sl SIAM J. Sci. Comput.} {\bf 42}(5)(2020), pp. B1193-B1226.
		

		\bibitem{jiang} {\sc D. Jiang, M. Mao, J. Li, and X. Deng},
		{\em An implicit parallel UGKS solver for flows covering various regimes}, {\sl Adv. Aerodyn. } {\bf
			1}(1) 8 (2019).
		
		
		\bibitem{mieussens} {\sc L. Mieussens},
		{\em Discrete-Velocity Models and Numerical Schemes for the
			Boltzmann-BGK Equtaion in Plane and Axisymmetric Geometries}, {\sl
			J. Comput. Phys.} {\bf 162} (2000), pp. 429-466.
		
		
		\bibitem{li} {\sc Z.H. Li and H.X. Zhang},
		{\em Gas-Kinetic Numerical Studies of Three-Dimensional Complex
			Flows on Spacecraft Re-Entry}, {\sl J. Comput. Phys.} {\bf 228}
		(2009), pp. 1116-1138.
		
		\bibitem{liu} {\sc C. Liu, Y. Zhu, K. Xu},
		{\em Unified gas-kinetic wave-particle methods I: Continuum and rarefied gas flow}, {\sl J. Comput. Phys.} {\bf 401}
		(2020), 108977.
		
		\bibitem{liu2014unified}
		S.~Liu, P.~Yu, K.~Xu, C.~Zhong, Unified gas-kinetic scheme for diatomic
		molecular simulations in all flow regimes, Journal of Computational Physics
		259 (2014) 96--113.
		
		
		\bibitem{ohwada} {\sc T. Ohwada},
		{\em Structure of normal shock waves: direct numerical analysis of
			the Boltzmann equation for hard-sphere molecules}, {\sl Phys. Fluids
			A} {\bf 5} (1993), pp. 217-234.
		
		
		
%		\bibitem{Pieraccini} {\sc S. Pieraccini and G. Puppo},
%		{\em Implicit-Explicit Schemes for BGK Kinetic Equations}, {\sl J.
%			Scientific Computing} {\bf 32} (2007), pp. 1-28.
		
		
%		\bibitem{pham} {\sc G.C. Pham-Van-Diep, D.A. Erwin, and E.P. Muntz},
%		{\em Nonequilibrium molecular motion in a hypersonic shock wave},
%		{\sl Science} {\bf 245} (1989), pp. 624.
		
		
		\bibitem{rykov} {\sc V. Rykov, V. Skobelkin},
		{\em Macroscopic description of the motions of a gas with rotational degrees of freedom}, {\sl Fluid Dyn.}
		{\bf 13} (1), (1978), 144-147.
		
		\bibitem{shakhov} {\sc E.M. Shakhov},
		{\em Generalization of the Krook kinetic Equation}, {\sl Fluid Dyn.}
		{\bf 3}, 95 (1968).
		
		
%		\bibitem{steinhilper} {\sc E.A. Steinhilper},
%		{\em Electron beam measurements of the shock wave structure: Part 1,
%			the inference of intermolecular potential from shock structure
%			experiments}, {\sl Ph.D. Thesis}, California Institute of Technology
%		(1972).
		
		
		\bibitem{xu-book} {\sc K. Xu},
		{\em Direct Modeling for Computational Fluid Dynamics: Construction and Application of Unified
			Gas-kinetic Scheme}, {\sl World Scientic, Singapore} (2015).
		
		\bibitem{xu} {\sc K. Xu},
		{\em A Gas-Kinetic BGK Scheme for the Navier-Stokes Equations and
			Its Connection with Artificial Dissipation and Godunov Method}, {\sl
			J. Comput. Phys.} {\bf 171} (2001), pp. 289-335.
	
	\bibitem{xu-tau} {\sc K. Xu},
	{\em Regularization of the Chapman–Enskog expansion and its description of shock structure}, {\sl
			Phys. Fluids},  {\bf 14} (4) (2002), pp. L17-20.
	
		\bibitem{xu-huang} {\sc K. Xu and J.C. Huang}, {\em A unified gas-kinetic scheme for continuum
			and rarefied flows}, {\sl J. Comput. Phys.} {\bf 229}, pp. 6715-6731 (2010).
		
		
		\bibitem{xu-c} {\sc X. Xu, Y. Chen, C. Liu, Z. Li, and K. Xu}, {\em Unified gas-kinetic wave-particle methods V: diatomic molecular flow},
		arXiv:2020.07195v1 [physics.comp-ph] 14 Oct. 2020.
		
		
		
		\bibitem{yang} {\sc J.Y. Yang and J.C. Huang},
		{\em Rarefied flow computations using nonlinear model Boltzmann
			equations}, {\sl J. Comput. Phys.} {\bf 120} (1995), pp. 323-339.

	\bibitem{yu} {\sc P.B. Yu},
		{\em A unified gas kinetic scheme for all Knudsen number flows}, {\sl PhD Thesis} (2013), Hong Kong University of Science and Technology.

		
		\bibitem{zhu} {\sc Y. Zhu, C. Liu, C. Zhong, K. Xu},
		{\em Unified gas-kinetic wave-particle methods. II. multiscale simulation on unstructured mesh}, {\sl Physics of Fluids} {\bf 31}(6) (2019) 067105.
		
	
		
	\end{thebibliography}
\end{document}